\documentclass{emulateapj}

\usepackage{natbib}
\usepackage{xspace}
\usepackage{hyperref}
\usepackage{graphicx}

\newcommand{\noprint}[1]{}
\newcommand{\figsetstart}{{\bf Fig. Set} }
\newcommand{\figsetend}{}

\newcommand{\figsetnum}[1]{{\bf #1.}}
\newcommand{\figsettitle}[1]{ {\bf #1} }

\begin{document}
\shorttitle{A sensitive spectral survey of interstellar features in
  the near-UV} \shortauthors{N.~H.~Bhatt \& J.~Cami} \title{A
  sensitive spectral survey of interstellar features in the near-UV
  [3050-3700\AA].}

\author{N.~H.~Bhatt\altaffilmark{1}}
\author{J.~Cami\altaffilmark{1,2}}
\affil{\altaffilmark{1}Department of Physics and Astronomy, The
  University of Western Ontario, London, ON N6A 3K7, Canada \\
\altaffilmark{2}SETI Institute, 189 Bernardo Avenue, Suite 100,
Mountain View, CA 94043, USA}  
\email{nbhatt3@uwo.ca;jcami@uwo.ca}
\slugcomment{Accepted for publication in \apjs}

\begin{abstract}

We present a comprehensive and sensitive unbiased survey of
interstellar features in the near-UV range (3050-3700 \AA). We
combined a large number of VLT/UVES archival observations of a sample
of highly reddened early type stars -- typical diffuse interstellar
band (DIB) targets -- and unreddened standards.  We stacked the
individual observations to obtain a reddened ``superspectrum'' in the
interstellar rest frame with a signal-to-noise (S/N) ratio exceeding
1500. We compared this to the analogous geocentric and stellar rest
frame superspectra as well as to an unreddened superspectrum to find
interstellar absorption features. We find 30 known features (11 atomic
and 19 molecular) and tentatively detect up to 7 new interstellar
absorption lines of unknown origin. Our survey is sensitive to narrow
and weak features; telluric residuals preclude us from detecting
broader features. For each sightline, we measured fundamental
parameters (radial velocities, line widths, and equivalent widths) of
the detected interstellar features. We also revisit upper limits for
the column densities of small, neutral polycyclic aromatic hydrocarbon
(PAH) molecules that have strong transitions in this wavelength range.
\keywords{ ISM: atoms -- ISM: molecules --line: identification --
  methods: observational -- stars: early-type}
\end{abstract}

%%%%% INTRODUCTION %%%%%%%%%%%%%%%%%%%%%%%
\section{Introduction}
\label{Sect:Intro}

The interstellar medium (ISM) is chemically rich: almost 200 molecular
carriers have been detected in space \citep[see
  e.g.][]{2001A&A...370L..49M, 2005JMoSt.742..215M,
  2013RvMP...85.1021T} as well as many atomic species. In addition,
there remains an unidentified class of gas phase molecular carriers
thought to be responsible for a series of over 400 interstellar
features - the diffuse interstellar bands (DIBs) - that are observed
in lines of sight that intersect interstellar clouds (see
\citealp{Herbig:1995,2006JMoSp.238....1S} for reviews; for DIB
surveys, see \citealp{Jenniskens:survey, Seathrun:survey,
  2008ApJ...680.1256H, 2009ApJ...705...32H}). The DIB carriers are
presumably a collection of stable, carbonaceous molecules that are
widespread and abundant in space. They are most likely related to
carbon chains, polycyclic aromatic hydrocarbons (PAHs) or fullerene
compounds.

Many molecular species -- including some of these proposed DIB
carriers -- could be in principle detected and identified in the UV
and the near-UV. For instance, laboratory measurements of small
neutral PAHs show strong transitions in the near-UV. Recently,
\citet{2011ApJ...728..154S} and \citet{2011A&A...530A..26G} therefore
compared their laboratory data to near-UV archival data of reddened
targets. No PAH features were reliably detected, but
\citet{2011A&A...530A..26G} found several other molecular absorption
lines in this range (due to CH, CH$^{+}$, NH, CN, and even OH$^{+}$).

However, while the ISM has been extensively studied in the optical and
infrared ranges, the near-UV remains relatively unexplored in
comparison, and thus our inventory of interstellar spectral features
may be incomplete.  Indeed, to the best of our knowledge, an unbiased
search for interstellar features and possible DIBs in this wavelength
range has not yet been presented in the literature.  This lack of
detailed studies reflects the challenges of observing in the near-UV
such as generally poor instrumental sensitivity and significant
telluric ozone absorption. Here, we present such an unbiased survey
using archival data from the UVES instrument mounted on the VLT. UVES
is one of the few instruments that has a good sensitivity in the
near-UV; we furthermore increase our sensitivity by co-adding a large
number of spectra.

This paper is organized as follows. We begin by detailing our data set
and processing steps in
\S\ref{Sec:observationaldata}. \S\ref{Sec:search} then describes the
methodology we used to search for interstellar features, our
sensitivity limits and results. We discuss our results in
\S~\ref{Sec:discussion}, and also revisit upper limits on the column
densities of small, neutral PAHs. We summarize our work in
\S~\ref{Sec:conclusion}.

%%%%% OBSERVATIONS AND DATA REDUCTION %%%%%%%%%%%%%%%%%%%%%%%
\section{Observations and data reduction}
\label{Sec:observationaldata}

Our goal is to search in the near--UV for interstellar lines, and
those are often weak and narrow. We thus require at the same time high
spectral resolution and good sensitivity. Very few instruments offer
both at the same time in the near-UV. The UVES instrument \citep{UVES}
mounted on the Very Large Telescope (VLT) is ideally suited since it
covers the range $\sim$3000\AA--11000\AA\ at high spectral
  resolution.  For the near-UV range of interest, we only need data
using the standard 346 setting (centered around 346 nm) on the blue
arm of the instrument, covering the range 3030\AA-3880\AA ~at a
  maximum resolving power of 82,000. The ESO archive contains a
large data set of UVES 346 observations (both raw and fully reduced
data) for many targets that are suitable for our study here, and we
thus exclusively use UVES archival data in this paper.

\subsection{Sample Selection}

The ideal targets for our survey are stars that sample a sight line
with significant amounts of interstellar material, that exhibit few
confusing stellar lines, and that are bright enough to have sufficient
flux in the near-UV. Reddened, early-type (O and B) stars are thus
ideal targets -- criteria that are also characteristic for DIB
targets. We searched the UVES archive for such targets, and not
surprisingly, a significant fraction of the data we use here stems
indeed from observational DIB programs or from programs studying
translucent clouds. Our study here includes 185 observations of 51
reddened targets (see Table~\ref{Table:astrodata}).

In addition to our reddened targets, we also searched for a large
number of unreddened comparison stars. Such standards are important to
better establish the nature (instrumental, telluric, stellar or
interstellar) of spectral features.  We found these targets by
searching SIMBAD for all stars with spectral type O or B with $B-V <
0.1$ and with at least 100 literature references.  Using this list of
$\sim$1000 targets, we searched the UVES archives and obtained 300
observations corresponding to 50 unreddened targets. From this
collection, we then visually inspected each observation, discarding
observations that showed a clear interstellar \ion{Na}{1} line, or
exhibited either possible reduction problems or instrumental
residuals. After this filtering, 166 observations corresponding to 39
unreddened targets remained.

Our final data set thus consists of 351 spectra, obtained with a
resolving power between 58,000--71,000 for the reddened targets,
  and 40,000--71,000 for the unreddened targets. The full list of
targets, along with basic data and the corresponding ESO programs are
listed in Table~\ref{Table:astrodata}.

\subsection{Data Reduction}

For most of our targets, we obtained fully reduced spectra from the
UVES-ESO reprocessed archive%
\footnote{{\texttt
    http://archive.eso.org/eso/eso\_archive\_adp.html}}.  However, for
some targets, clear artefacts were present in these standard pipeline
products (due to e.g. incorrect order merging). For these targets, we
obtained the raw data%
\footnote{{\texttt
    http://archive.eso.org/wdb/wdb/eso/eso\_archive\_main/query}} and
carried out the data reduction ourselves. We similarly retrieved the
raw data from the archive for those targets for which fully reduced
data were not yet available in the reprocessed archive.

We processed the raw data using the Gasgano\footnote{{\texttt
    http://www.eso.org/sci/software/gasgano/}} file organizer with the
UVES pipeline (version 5.0.9), and followed the \verb+uves_obs_scired+
recipe to reduce the raw observations. This recipe performs bias and
dark frame subtraction, flat-fielding, wavelength calibration,
automated order detection, background subtraction, optimal extraction,
and optimal merging. We retrieved the necessary observation-specific
calibration files (bias, dark, flat and arc lamp frames) from the ESO
archive. Other calibration files (catalog of arc lines, arc lamp
frame, atmospheric extinction table, standard star, format check
frame) are provided with Gasgano. The final product is a full,
calibrated spectrum, similar to what is provided in the reprocessed
archive.

A visual inspection of all reduced spectra revealed systematic
artefacts in some observations, showing up as repeating undulations
with a period of one echelle order. This issue is well
documented
\footnote{\texttt
  http://www.eso.org/observing/dfo/quality/reproUVES/processing.html},
but is not easily corrected. In the most severe cases, we excluded the
spectra from our study. We kept some of the mildly affected spectra,
reasoning that the artefacts are much broader than the typical
features we will be searching for and thus should not interfere much
with our goals. The signal-to-noise ratio is variable across the spectrum. The
spectra are quite noisy at the shortest wavelengths (around 3040\AA)
due to the low sensitivity of the instrument at these wavelengths, but
the quality improves at longer wavelengths; Table~\ref{Table:astrodata}
gives some indication of how the quality varies across our sample.
 
\subsection{Further Processing}
\label{Sec:FutherProcessing}

\begin{figure}[t]
\centering
\resizebox{\hsize}{!}{\includegraphics{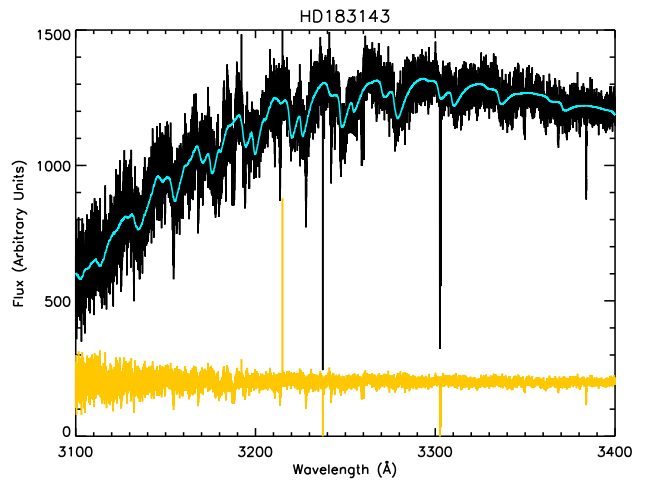}}
\caption{\label{Fig:preprocess-ozone}An example of the telluric ozone
  correction, where we have divided the UVES spectrum of HD~183143
  (top; black) by the best fit telluric ozone spectrum (top; cyan),
  resulting in the corrected spectrum (bottom; gold; shown here
  $\times$200 for clarity).  }
\end{figure}

\begin{figure}[t]
\resizebox{\hsize}{!}{\includegraphics{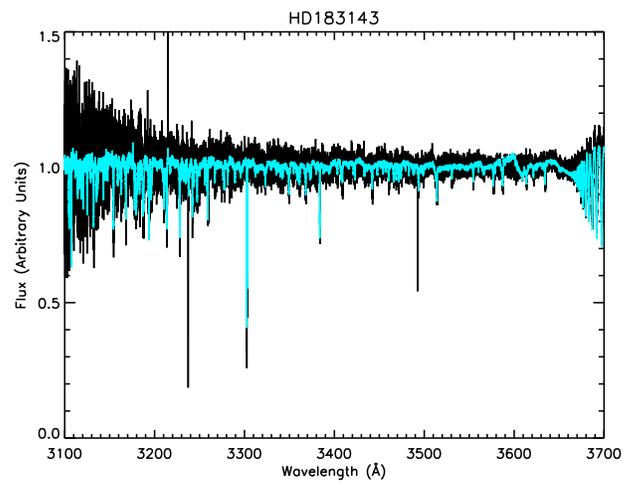}}
\caption{\label{fig:preprocess-add}An example of the gain in quality
  obtained from co-adding observations. A single, processed spectrum
  of HD~183143 is shown in black, and the co-added spectrum (including
   four observations) is shown in cyan. }
\end{figure}

We carried out some additional processing steps to facilitate further
analysis and improve the sensitivity of our search. First, we flagged
bad pixels and cosmic ray hits, and replaced the flux values at these
points with interpolated values while assigning a large uncertainty. 

Next, we corrected our spectra for the strong contamination due to
telluric ozone in the near-UV.  In our spectra, absorption in the
Huggins band is clearly present in the range 3100-3400 \AA\ (see
Fig.~\ref{Fig:preprocess-ozone}).  Ideally, one would divide the
  spectrum of a reddened target by the spectrum of an unreddened,
  broad-lined reference star of the same spectral type observed at a
  similar airmass and close in time to the target. Good pairwise
  reference observations however are not available in the archive for
  most of our targets. Furthermore, dividing by a standard decreases
  the S/N in the bluest parts of the spectrum where sensitivity is
  already low. We therefore opted for an alternative approach to
  remove the ozone signature, similar to \cite{Herbig:C60}. We used
O$_3$ absorption cross sections from \citet{Burrows:ozone} and
determined the ozone column density for each observation that provides
the best match to the spectrum in that range. We then divided out the
ozone spectrum from the observations. As can be seen from
Fig.~\ref{Fig:preprocess-ozone}, this removes all of the broad
structure visible in the spectrum. However, since the cross sections
are sampled at a much lower spectral resolution than our observations,
narrow telluric features may still be left in the spectra (see
\S~\ref{Sec:limitations}). We also compared our spectra to the
absorption cross-sections of several other atmospheric species
(e.g. BrO, ClO$_2$, H$_2$CO, and
SO$_2$)\footnote{http://hitran.iao.ru/xsections} but did not find
evidence for any of them in our spectra. Note that narrow
  telluric residuals remain, but will not pose insurmountable
  problems for our work (see Sect.~\ref{Sec:search}).

Next, we determined the broad-band continuum for each our spectra by
fitting a cubic spline through a small number of anchor points that
were iteratively determined to best reproduce the observed stellar
continuum. The resulting curve reproduces all broad-band spectral
variations, including many (broad) artefacts remaining from the data
reduction process. Our routine has some difficulties near the Balmer
discontinuity, resulting in artefacts in the spectrum. However, these
will not affect our further analysis much. After inspection, we
normalized our observations to this continuum. 

We then applied a heliocentric correction to all individual
observations, and determined the signal-to-noise ratio of each
spectrum from the mean flux and standard deviation in a wavelength
range devoid of obvious spectral features (3450\AA-3452\AA). For
  each target, we then co-added the individual spectra by calculating
  the weighted mean flux at each wavelength where the weights are
  inversely proportional to the variance in the data. For the
  remainder of this paper, we will refer to this weighted average
  spectrum of a target as the {\em optimal
    spectrum}. Fig.~\ref{fig:preprocess-add} illustrates the dramatic
gain in quality that results from this operation. For our survey
described below, we thus start from a single, high-quality optimal
spectrum for each of our 90 reddened and unreddened targets.

%%%%% SEARCH FOR INTERSTELLAR FEATURES %%%%%%%%%%%%%%%%%%%%%%%
\section{A search for interstellar features}
\label{Sec:search}

\subsection{Methods}

\begin{figure}[t]
\resizebox{\hsize}{!}{\includegraphics{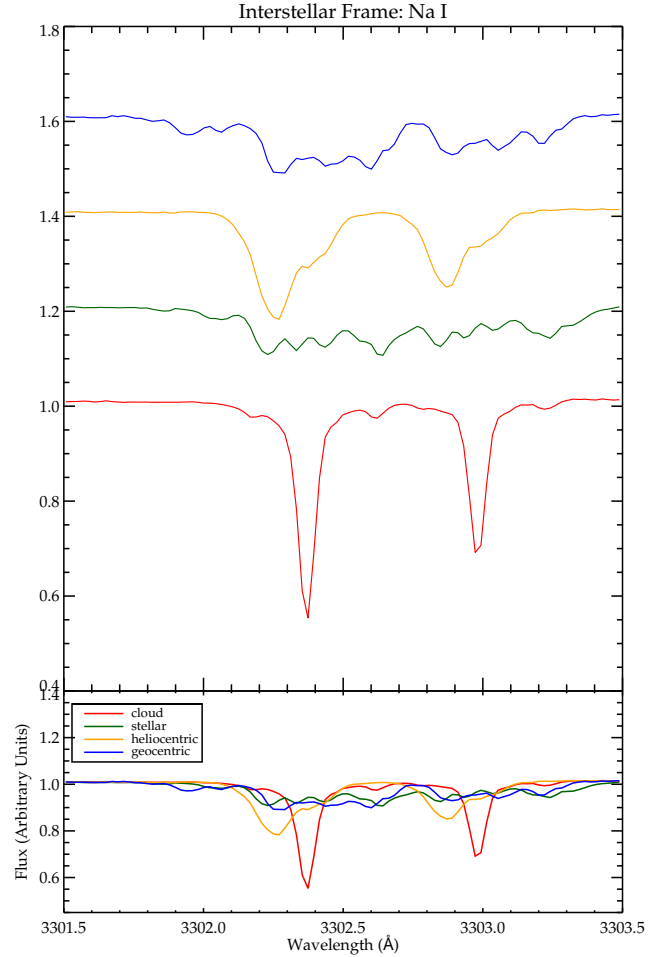}}
\caption{\label{fig:comparison-method-NaI} The four superspectra in
  the spectral region around the \ion{Na}{1}
  3302.368/3302.987\AA\ doublet for respectively the geocentric (top;
  blue), heliocentric (yellow), stellar (green) and interstellar
  (bottom; red) rest frame. The bottom panel shows all four spectra on
  top of each other. }
\end{figure}

\begin{figure}[t]
\resizebox{\hsize}{!}{\includegraphics{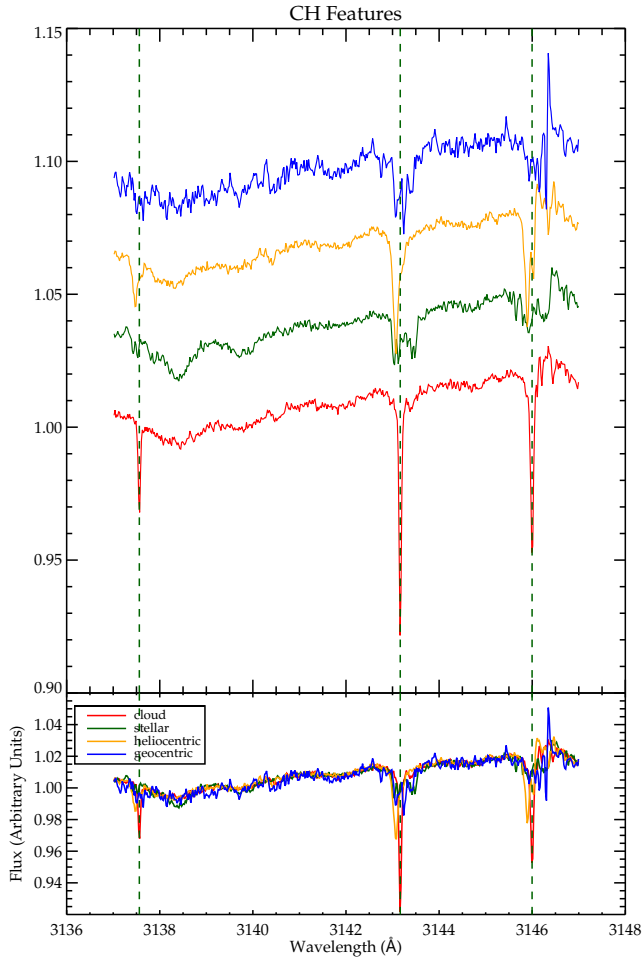}}
\caption{\label{fig:known_CH}The geocentric (top; blue), heliocentric
  (yellow), stellar (green) and interstellar (bottom; red)
  superspectra between 3137\AA\ and 3148\AA. A triplet of CH features
  shows up prominently in our interstellar superspectrum. These lines
  appeared moderately strong in about half of the reddened
  sightlines. }
\end{figure}

Interstellar spectral features can have different appearances. For
lines of sight that intersect only one interstellar cloud, atomic
(ionic) and simple molecular lines are generally very narrow since
their intrinsic line width is determined primarily by the velocity dispersion in
the cloud. Such lines are generally easy to detect and recognize.
DIBs on the other hand have a wide range of intrinsic widths and
profiles \citep[see e.g.][]{1988A&A...203..134W, 1996A&A...307L..25E,
  1996MNRAS.283L.105K, 1997ApJ...477..209K, 2001MNRAS.323..293G,
  2002ApJ...578..877S, 2002A&A...396..987G, 2008ApJ...682.1076G}. 

Several methods can be used to search for interstellar
features. Observations of spectroscopic binaries at different orbital
phases provide a clear way to separate stellar from interstellar lines
owing to the stationary nature of the latter lines \cite[see e.g.][for
  a recent example]{2008ApJ...680.1256H}. While powerful, this method
can only be used for lines of sight where a suitable binary is
available. DIB surveys have therefore often reverted to using a
pair-wise comparison between reddened targets and unreddened standards
of the same spectral type to distinguish stellar from interstellar
lines \citep[see e.g.][]{Herbig:1975, Jenniskens:survey,
  Seathrun:survey, 2009ApJ...705...32H}. However, even with a good
spectral match, some ambiguities often remain in recognizing
(especially broad) interstellar lines \cite[see
  e.g.][]{2009ApJ...705...32H}.

Here, we tried out a different approach that is based on having a
large sample of reddened targets. The basic premise is that stellar
and interstellar radial velocities are generally quite different
  for each of our targets, and are either known and/or easy to
determine. By shifting all of our target spectra to their respective
interstellar cloud rest frame, interstellar features line up, while
stellar lines are spread out; co-adding will thus result in stellar
lines that are broadened and weakened, but the interstellar lines
should be fairly prominent (and relatively narrow). Similarly, when
shifting to the stellar rest frame and co-adding, stellar lines are
emphasized while interstellar lines are smeared out. A comparison
between these two co-added {\em ``superspectra''} should thus -- in
principle -- readily reveal all interstellar features. Additionally,
the co-added superspectrum has a much higher signal-to-noise ratio
which allows to detect even weak features (provided that they are
weakly present in a majority of the targets). An important caveat
though is that most of our reddened targets are not single cloud lines
of sight, i.e. their interstellar lines show more than one component
at different radial velocities. Some interstellar features may not be
present in all components.

In practice, we first determined stellar radial velocities for all of
our targets from cross-correlating the observed spectrum with a simple
synthetic spectrum comprised of only hydrogen and helium lines. To
determine the interstellar cloud velocities, we first note that all
(reddened) targets clearly show the narrow absorption lines in
the \ion{Na}{1} doublet at 3302\AA. In all cases, we have measured the
radial velocity corresponding to the deepest absorption in this
\ion{Na}{1} doublet. Table~\ref{Table:astrodata} lists all radial
velocities we measured in this way. Where available, we have
  compared our results to literature values and found good
  agreement. 

We then created four different superspectra (see
e.g. Fig.~\ref{fig:comparison-method-NaI}), and refer to them by the
radial velocity rest frame to which the individual spectra were
shifted before co-adding. The geocentric superspectrum corresponds to
no shifts at all, and is useful to recognize telluric features. The
heliocentric superspectrum is the co-add after applying just a
heliocentric correction to the spectra. We then shifted all spectra to
their stellar rest frame, and the resulting co-add is the stellar
superspectrum. Finally, the interstellar superspectrum results from
co-adding the spectra shifted to the interstellar rest frame.  Note
that each of the superspectra is a weighted mean, where the
weights are again inversely proportional to the variance of the
individual optimal spectra. 

To illustrate our method, we first present in
Fig.~\ref{fig:comparison-method-NaI} the four superspectra around the
\ion{Na}{1} doublet near 3302\AA. As expected, the \ion{Na}{1} lines
are sharp and well-defined in the interstellar superspectrum. Note
that the superspectrum even records a smaller secondary peak
originating from the fact that most lines of sight have multiple
clouds in the line of sight. Fig.~\ref{fig:known_CH} shows
the same superspectra around a triplet of known interstellar CH features
at 3137.530\AA, 3143.150\AA~and 3146.010\AA. Also these lines are much better
defined in the interstellar superspectrum, thus establishing their
interstellar nature. This forms the basis of our search for
interstellar features in the near-UV.

\subsection{Sensitivity \& Limitations}
\label{Sec:limitations}

\begin{figure*}[t]
\resizebox{\hsize}{!}{\includegraphics{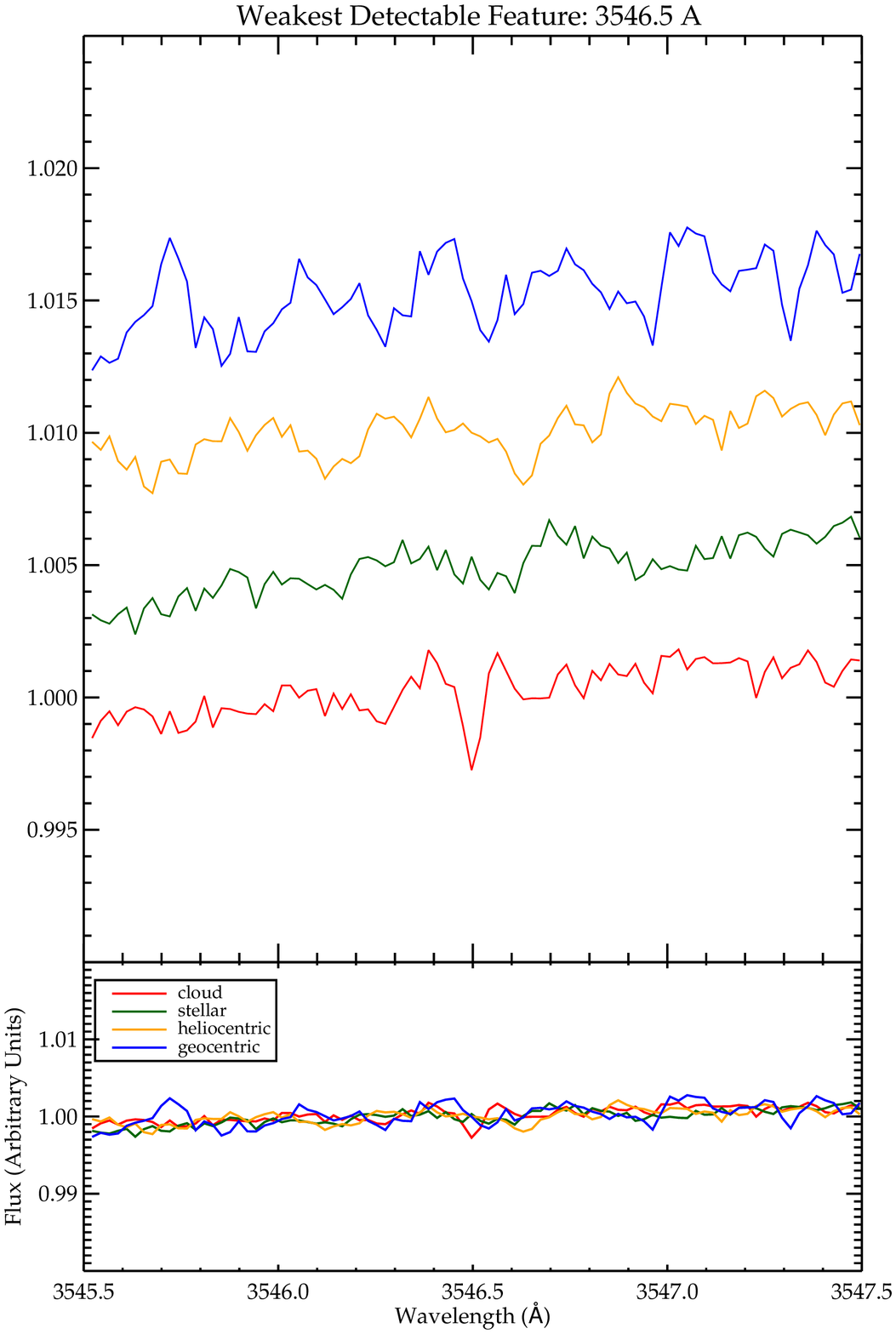}
                     \includegraphics{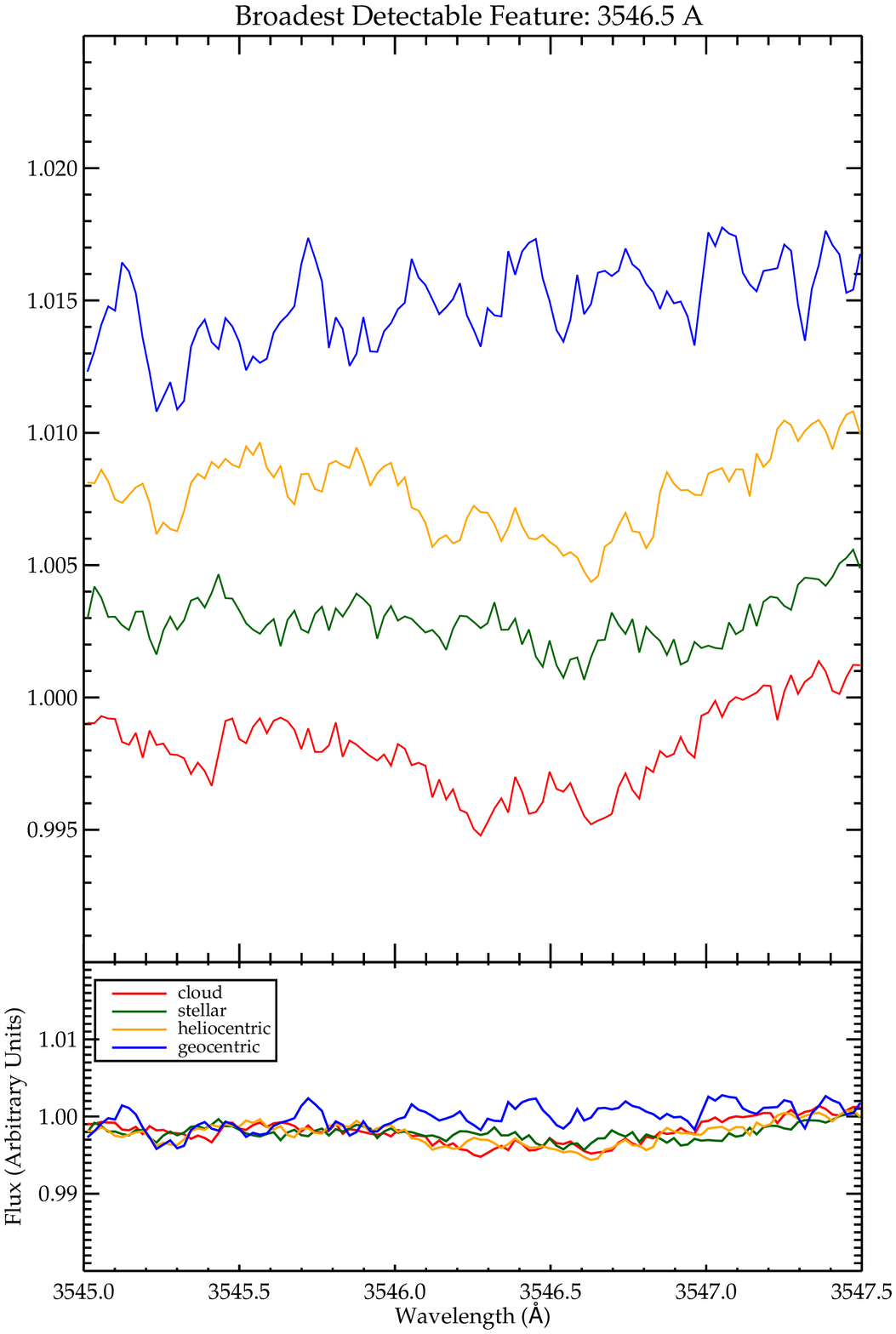}}
\caption{\label{Fig:featurelimits} The superspectra after
    introducing an artificial interstellar feature in each of the
    target optimal spectra (at an interstellar rest wavelength of
    $\lambda$=3546.500\AA).  Both panels show the geocentric
  (top; blue), heliocentric (yellow), stellar (green) and interstellar
  (bottom; red) superspectra.  Note that the geocentric superspectrum
  does not contain the artificially introduced features and is shown
  only for comparison. ({\em Left:}) the artificial feature is an
  unresolved line with an equivalent width of $W=0.3$m\AA. This is
  about the weakest unresolved feature we can detect. ({\em Right:})
  the artificial feature now has a FWHM of 0.8\AA\ and a depth of
  about 0.5\% (corresponding to an equivalent width of $W\approx
  6$m\AA). This is about the broadest weak feature that we can
  reliably detect as an interstellar feature. }
\end{figure*}

Throughout our entire wavelength range, the S/N of the interstellar
superspectrum is typically $\ge$1500. Thus, the 3$\sigma$ detection
limit for an unresolved line (FWHM $\sim$45m\AA) is technically
$W\approx 0.1$m\AA. However, we found that this is an optimistic
estimate of our sensitivity, since a detection stems from a comparison
to the other superspectra, and those have different noise
characteristics (see below). As a test, we introduced a synthetic
feature in the optimal spectra of each of our targets and tried to
detect it in the superspectrum. We found that a more realistic
estimate of our detection limit is $W\ga 0.3$m\AA\ (as illustrated in
Fig.~\ref{Fig:featurelimits}).

It is clear that the apparent S/N in the various superspectra (as
measured from the variance in the flux in a region without clear
spectral lines) is quite different, even though they result from
co-adding the same optimal target spectra (see
Fig.~\ref{Fig:featurelimits}). The S/N is systematically lowest (by a
large margin) for the geocentric superspectrum. This is not indicative
of the quality of the spectra, but must be the consequence of having a
large number of weak and narrow telluric features in our spectra that
get smeared out in the other superspectra. While it is reassuring to
see that very little is left of these features in the interstellar
superspectrum, it turns out that chance superpositions sometimes
result in weak features in the other superspectra. This is further
discussed in Sect.~\ref{Sec:results}.

A clear limitation of our method is that we are not sensitive to broad
(and shallow) interstellar features. Indeed, for broad lines, there is
not enough contrast between the interstellar and stellar superspectra
to reliably detect such features or to tell them apart from what could
be stellar lines. To get a sense of this limit, we once more
introduced a feature in each of the optimal spectra, with a depth of
0.5\% and changing the FWHM of the feature until we could not reliably
identify it as an interstellar feature anymore. We found that this
limit is at a FWHM of about 0.8\AA, corresponding to an equivalent
width of $W\approx 6$m\AA. This is thus the broadest weak interstellar
feature we could hope to detect with our method (see
Fig.~\ref{Fig:featurelimits}).

In the process of searching for interstellar lines described below, we
also considered some variations on our strategy to improve the
detection probability. Sometimes, it is advantageous to not include
all lines of sight in the superspectrum. Indeed, if a feature only
appears in a small number of sightlines, the feature would be dampened
in the superspectrum by the larger number of spectra where the feature
is absent. Thus, such a feature may be more easily detected by
excluding these latter targets and making a superspectrum of a smaller
subset of spectra. This then results in a larger contrast in the
feature, which can sometimes more than compensate for the lower S/N of
the resulting co-add. However, in practice we detected all
  features in the superspectra comprised of all 51 reddened targets
  although we found it helpful to consider well-defined subsets of
  targets (e.g. the most or least reddened) for distinguishing between
  artefacts and potential interstellar features.

  Finally, we also compared the reddened superspectrum to 
the corresponding unreddened superspectrum (see
e.g. Fig.~\ref{Fig: RvsUR}). Note that stellar lines between the two
do not match, since the spectral types of the contributing targets in
both cases is different. However, it is often an insightful comparison
that can give more confidence that weak features are truly
interstellar lines.

\subsection{Detections}
\label{Sec:results}

\begin{deluxetable}{ll}
\tablecaption{\label{tab:features}Interstellar atomic, molecular and
  unidentified features in the near-UV.}
\tablehead{\colhead{Species} & \colhead{$\rm Wavelength (\AA,air)$ }}

\startdata

\ion{Cr}{1} & 3578.683\tablenotemark{a}, 3593.482\tablenotemark{a} \\
\ion{Fe}{1} & 3440.606\tablenotemark{a}, 3679.913\tablenotemark{a} \\
%\ion{K}{1}  & 3217.154\tablenotemark{a}           \\
\ion{Na}{1} & 3302.368\tablenotemark{a}, 3302.987\tablenotemark{a} \\
\ion{Ti}{2} & 3066.354\tablenotemark{a}, 3072.984\tablenotemark{a}, %
              3229.199\tablenotemark{a}, \\
            & 3241.994\tablenotemark{a}, 3383.768\tablenotemark{a}\\ 
\multicolumn{2}{c}{ }\\
CH          & 3137.530\tablenotemark{e}, 3143.150\tablenotemark{e}, 
              3146.010\tablenotemark{e}, \\
            & 3627.403\tablenotemark{d}, 3633.289\tablenotemark{d}, 
              3636.222\tablenotemark{d} \\  
CH$^+$      & 3447.075\tablenotemark{g}, 3579.020\tablenotemark{b}  \\
CN          & 3579.453\tablenotemark{f}, 3579.963\tablenotemark{f}, 
              3580.937\tablenotemark{f} \\
NH          & 3353.924\tablenotemark{b}, 3358.053\tablenotemark{b} \\
OH          & 3072.010\tablenotemark{c},  3072.064\tablenotemark{c}, 
              3078.440\tablenotemark{c},\\
            &   3078.472\tablenotemark{c}, 3081.664\tablenotemark{c}\\ 
OH$^+$      & 3583.769\tablenotemark{b} \\
Unident     &  3147.985\tablenotemark{*}, 3247.578\tablenotemark{*}, \\
            & 3346.968\tablenotemark{*}, 3362.942\tablenotemark{*}, \\
            &3572.656\tablenotemark{*} \\
\tablerefs{
(a) Peter van Hoof line list; 
(b) \citet{2011A&A...530A..26G}; 
(c) \citet{1996ApJ...465L..57F}; 
(d) \citet{2011AN....332..167W}; 
(e) \citet{1977ApJ...213..394C}; 
(f) \citet{1989ApJ...343L...1M}; 
(g) \citet{1960APJ...131....1D}; 
(*) this work.}

\enddata
\end{deluxetable}

To search for interstellar features, we started by comparing the
interstellar to the stellar superspectrum. We stepped through the
entire wavelength range in steps of 5\AA\ (see
Figs.~\ref{Fig:superspec}). We
visually inspected apparent absorption features, and marked them as
interstellar line candidates if they were detectable above the noise,
and if they appeared better defined in the interstellar superspectrum
compared to the stellar one. This resulted in 106 features (many of
them weak) that we further investigated.

We then compared the wavelengths of our features to those of known
atomic (ionic) transitions for various elements. For each of the
detected features, we searched for all possible transitions
(originating from the electronic ground state) that fall within
0.1\AA\ of the observed wavelength, using Peter van Hoof's
database\footnote{\url{http://www.pa.uky.edu/~peter/atomic/}}. This
way, we found 11 matching transitions due to 4 ions: \ion{Cr}{1},
\ion{Fe}{1}, \ion{Na}{1}, and \ion{Ti}{2} (see
Table~\ref{tab:features}).

We also searched for correspondences with known molecular features in
this wavelength range. To the best of our knowledge, no database with
molecular transitions in this wavelength range exists, and thus we
relied this time on various literature sources. We identified 19 of
our features with transitions due to 6 molecular species: CH, {\rm
  CH$^{+}$}, CN, NH, OH, and {\rm OH$^{+}$} (see
Table~\ref{tab:features}).

\begin{figure}[t]
\resizebox{\hsize}{!}{%
 \includegraphics{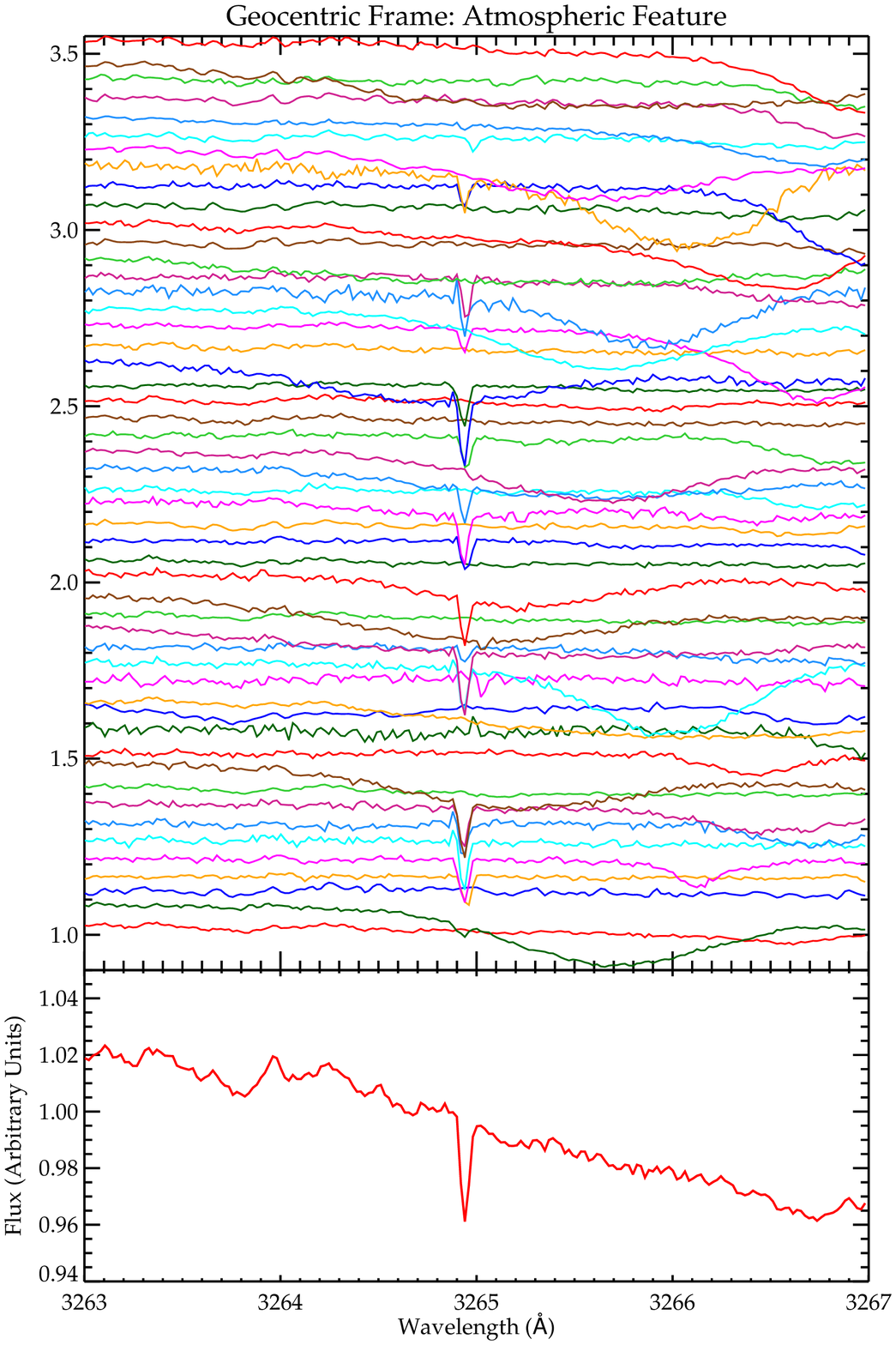}  
 \includegraphics{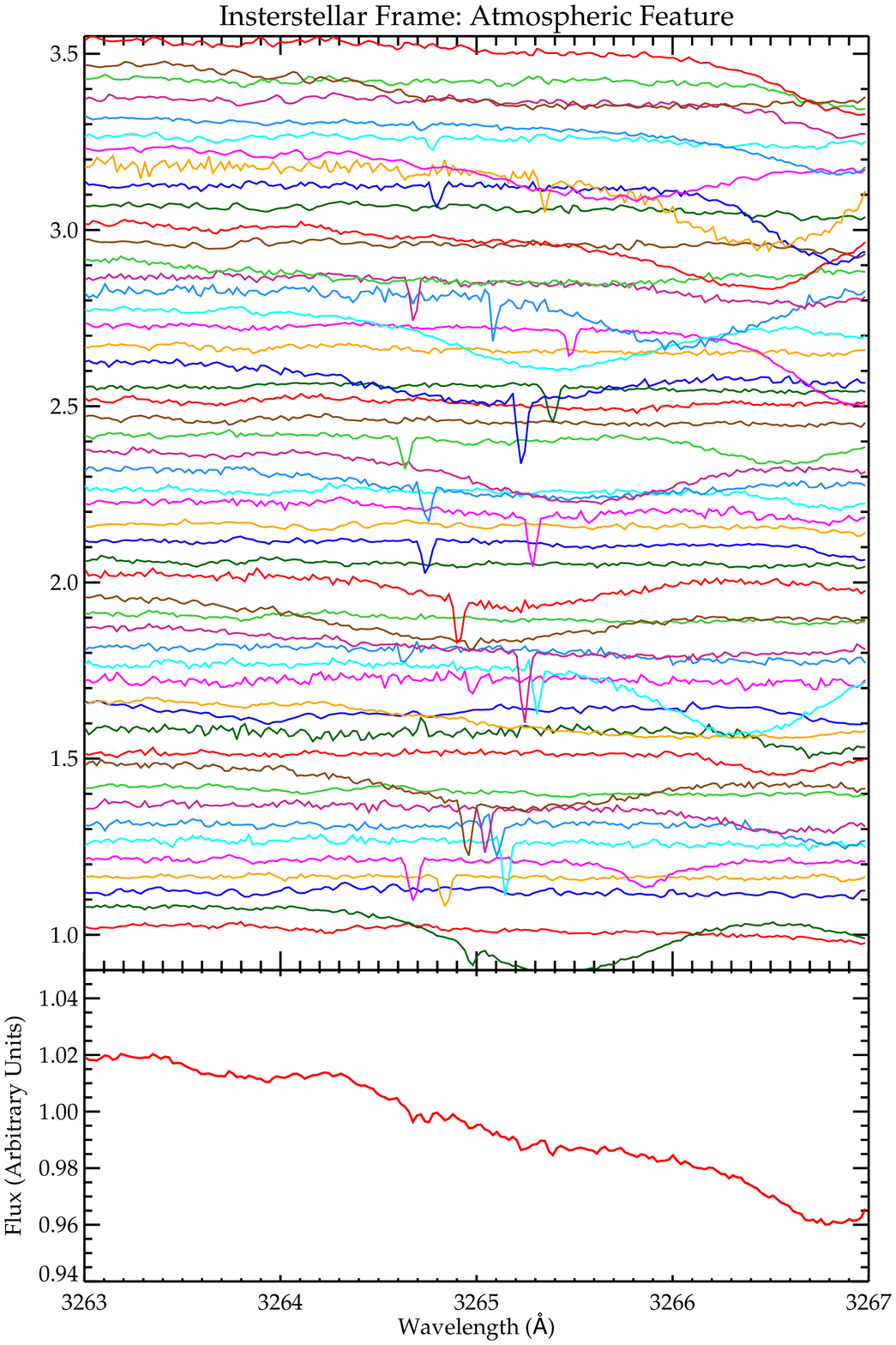}}
\caption{\label{Fig:atmospheric}({\em Left:}) A feature clearly aligns
  around 3265\AA\ in the optimal, geocentric spectra for all
  targets. Note that the varying relative strength suggests an
  atmospheric rather than an instrumental origin in this case. ({\em
    Right:}) In the interstellar rest frame, the feature appears at
  different wavelengths, and thus is greatly washed out in the
  interstellar superspectrum. However, chance alignments of
  interstellar clouds still result in very weak features. }
\end{figure}

We investigated the nature of the remaining 76 features in more detail
by comparing all of the superspectra, but also by studying the optimal
spectra of individual targets. We conclude that the majority of these
features are artefacts that are due to atmospheric and/or instrumental
issues. An example is shown in Fig.~\ref{Fig:atmospheric} where a
telluric feature is clearly present, but is not entirely washed out in
the interstellar superspectrum. In other cases, the origins of the
features are less clear, and sometimes several neighbouring weak
atmospheric features conspire to create features in the interstellar
superspectrum, making it even harder to establish the nature of
resulting absorption features.

Of the 76 remaining tentative features, we established that 71 are
certainly {\em not} interstellar features. The 5 remaining features
could possibly be very weak interstellar lines -- they all are near
our sensitivity limit. In investigating these features, we found 2
more very weak possible interstellar lines. 

\begin{figure*}[t]
\resizebox{\hsize}{!}{\includegraphics{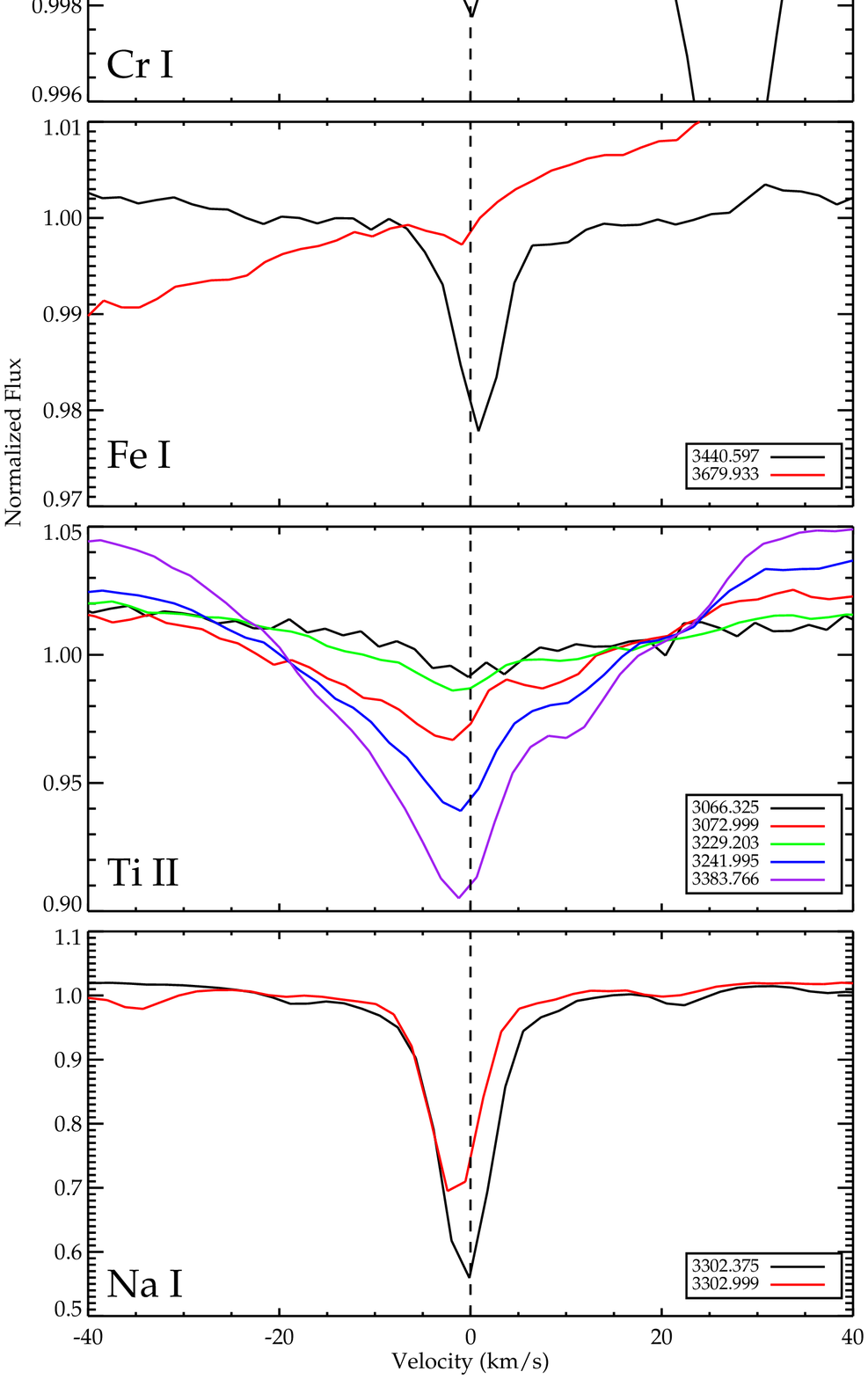}
                      \includegraphics{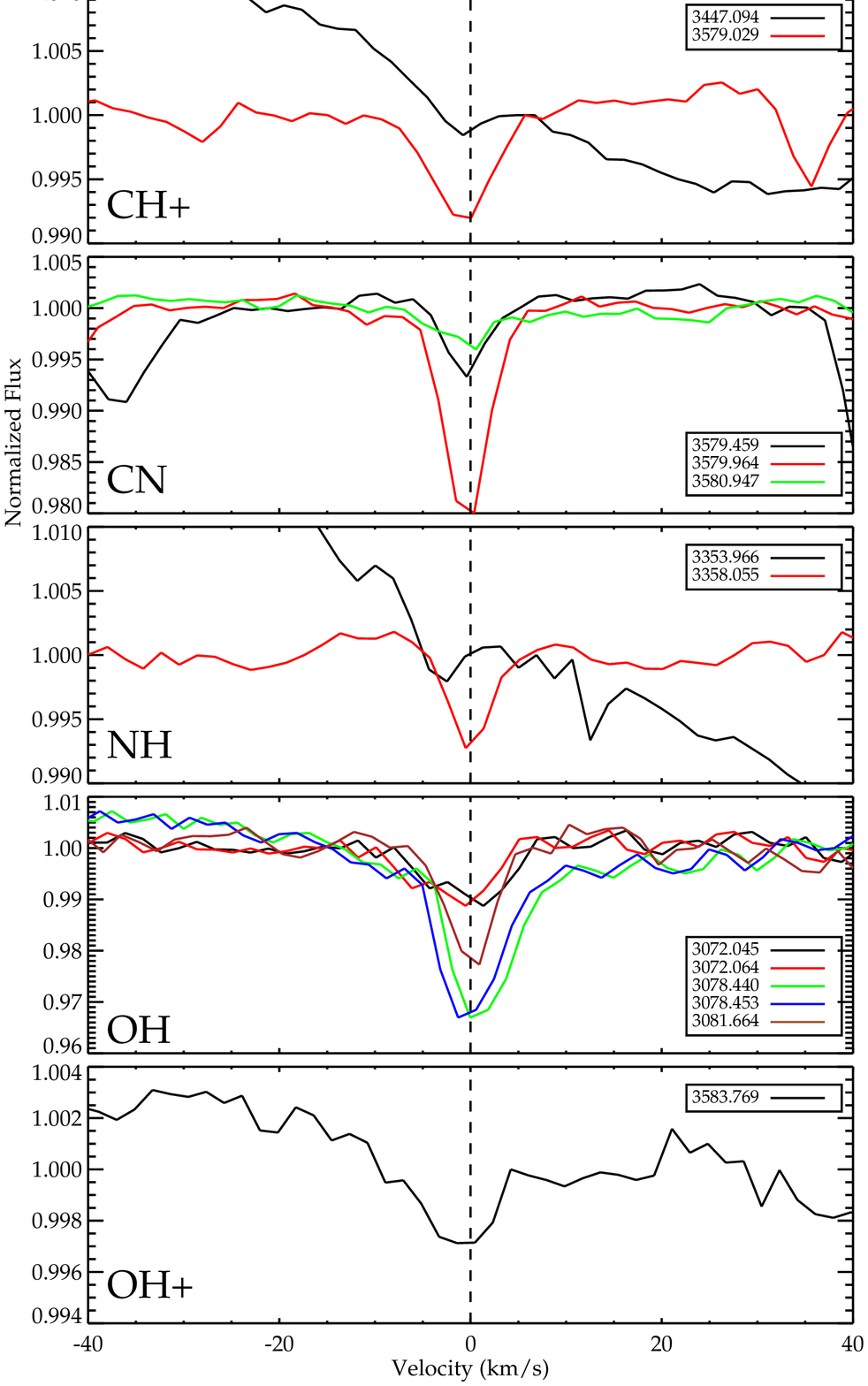}}
   \caption{\label{Fig:features_velspace}An overview of each
     identified interstellar feature in our superspectra: atomic {\em
       (left)} and molecular {\em (right)}.  The velocity scale is based on
     the interstellar superspectrum, that is, it is relative to the deepest
     component of the 3302  \ion{Na}{1} doublet.  Note that the OH lines are
     blends of several transitions. }
\end{figure*}

Thus, our careful search for interstellar features in the near-UV
results in 11 known atomic lines, 19 known molecular lines, and in the
best possible case 7 weak unidentified lines (see
Table~\ref{tab:features}). The resulting superspectra for each of the
features with identifications is shown in
Fig.~\ref{Fig:features_velspace}, and over the entire spectral range
in Figs.~\ref{Fig:superspec}; we discuss these features in some more
detail below.

As an independent check, we have also  created an additional
  superspectrum in the interstellar rest frame and a corresponding
  superspectrum using unreddened comparison stars using only those
  reddened stars for which we have good matching unreddened comparison
  stars. This subset is comprised of 26 pairs of reddened and
  unreddened targets; however, only 13 of the unreddened targets are
  unique. The resulting superspectra are shown in Fig~\ref{Fig:
    RvsUR}. A comparison between these two superspectra should also
reveal the interstellar lines quite readily, and could potentially
also turn up broader features.  We visually compared the reddened and
unreddened superspectra on a few different scales. Searching on
50\AA\ wide segments, we did not detect any moderate strength, broad
features (i.e. comparable to stellar lines). More detailed searches on
finer scales (20\AA, 5\AA\ segments) turned up all the features we
flagged before, and some more atmospheric features. These searches
suggest that there are no obvious broad interstellar lines (i.e. that
not blended with stellar lines; see Figs.~\ref{Fig: RvsUR}).

 It is reassuring that the detections found by comparing
 superspectra in various rest frames is in agreement with the more
 commonly used method of comparing reddened targets to unreddened
 standards. This suggests that comparing rest frames may be a viable
 alternative to the standard method when searching for interstellar
 lines. Our method has two clear advantages: it does not require
 unreddened standards, and it results in a higher sensitivity than a
 pairwise comparison from the gain in S/N when co-adding the optimal
 spectra. However, it requires a large number of observations to work
 properly, and is not sensitive to broad features. At the same time,
 it takes a significant amount of work to establish the true nature
 of detected features. Thus, a pairwise comparison of matching
 target/standard pairs may still be the best method to use if high
 quality observations of good spectral matches are available; in
 other cases, the rest frame comparison method may be an appealing
 alternative.

\subsection{Detected atomic lines}
\label{sec:detected_atomic_lines}

{\bf \ion{Na}{1}} -- Na is a well-known interstellar gas component
that is commonly detected in the well-known Na D lines at optical
wavelengths. For many lines of sight, these D lines are very
saturated. The Na U doublet lines near 3302\AA\ are generally less
saturated due to weaker transition probabilities
\citep{1974A&A....32....1D} -- they have 1.4\% the oscillator strength
of the Na D lines \citep{2002MNRAS.333..561K}. In conjunction with the
Na D doublet, these sodium lines can therefore be used to yield more
reliable Na column densities \citep{1980MNRAS.193..849B}. 

The Na U doublet is seen in all of our reddened targets, and even
weakly in several of our unreddened standards. In many cases, Doppler
splitting reveals 2 or more components. These lines are thus certainly
useful to determine Na column densities in all our lines of
sight. \\

{\bf \ion{Ti}{2}} -- We detect several lines of interstellar
\ion{Ti}{2} in most of our targets. In our wavelength range, there are
six possible transitions for interstellar \ion{Ti}{2} (3057.403,
3066.354, 3072.984, 3229.199, 3241.994 and 3383.768\AA). All but the
first line have been previously described \citep[see
  e.g.][]{1977ApJ...213..394C, 1996ApJ...463..181F,
  1996ApJ...465L..57F, 2010MNRAS.404.1321W}. In the UVES spectra, the
region around 3050\AA\ is furthermore very noisy, and thus we cannot
detect the first transition in our data set either. In our spectra,
the 3384\AA\ resonance line is the most notable.  It is commonly found
in the ISM, and is useful for estimating electron densities
\citep{1978ApJS...36..115S}. \\

{\bf \ion{Cr}{1}} -- In our wavelength range, there are three neutral
chromium transitions ($\lambda\lambda$ 3578.683, 3593.482,
3605.322). \citet{1990ApJ...349...91M} detected the first of these
weak features in $\zeta$ Oph with S/N=1250; all three lines are
detected toward HD~72127B \citep{2008MNRAS.388..323W}. In our
interstellar superspectrum, we detect the first two lines.  We also
detect these \ion{Cr}{1} lines in several targets. The \ion{Cr}{1} is
certainly a trace species; we find that our measured equivalent widths
for the strongest of the bands, 3578.683\AA, is about 1m\AA\ in our
superspectrum -- this is within the range determined by
\citet{2008MNRAS.388..323W} and \citet{1990ApJ...349...91M}.  It is
reasonable that we do not detect the 3605.322\AA~line since, of the
three features, it has the lowest oscillator strength
\citep{2003ApJS..149..205M}. \\

{\bf \ion{Fe}{1}} -- There are three possible Fe I transitions in
  our wavelength range: $\lambda\lambda$ 3440.606, 3649.302,
3679.913; of these, we detect the first and the third. These
transitions occur in a handful of sightlines.  The 3440.606\AA~feature
has been detected in HD~72127B, and QSO HE0001-2340
\citep{2008MNRAS.388..323W, 2007A&A...470..523D}. The non-detection of
the 3649.302\AA~line is consistent with its low absorption oscillator
strength -- it is a few orders of magnitude smaller than the detected
lines \citep{2006JPCRD..35.1669F}.\\

In addition to these detected lines, we also detected a very weak
feature at 3217.154\AA. This is right at the rest wavelength for the
$4s_{1/2}-7p_{3/2}$ transition of \ion{K}{1}, and we considered the
possibility that indeed this feature could be identified with
\ion{K}{1}. However, there are several more \ion{K}{1} transitions in
this wavelength range that should be stronger, but that we do not
detect. Moreover, the oscillator strength for this transition is far
too low \citep[$\sim10^{-4}$; ][]{2012PhRvA..86e2517N} to cause
  measurable absorption. We thus conclude that this is not a real
\ion{K}{1} feature.

Finally, we also searched for the 3130.42\AA\ and 3131.06\AA\ features
of interstellar \ion{Be}{1} \citep[see e.g.][]{1977ApJ...213..394C,
  1978ApJ...222..863M, 1985PASP...97...37B}; we do not detect either
of these transitions in our interstellar superspectrum nor in any
individual line of sight. Similarly, we could not detect the
\ion{Ni}{1} feature at 3369.565\AA\ \citep[see
  e.g.][]{1991AJ....101.1429M, 2008MNRAS.388..323W}.  Additionally,
like \citet{2008MNRAS.388..323W}, we did not detect \ion{Al}{1} 
  or \ion{Ti}{1} lines which could appear in our wavelength range.

\subsection{Detected molecular lines}

{\bf CH} -- Our spectra show strong interstellar CH lines; all 6 known
transitions in this spectral range are detected. A striking example of
the {\em C-X} system of CH is displayed in Fig.~\ref{fig:known_CH}.
\citet{2011AN....332..167W} report detections of the {\em B-X} system
around 3627\AA\ in 13 targets. For the targets we have in common, we
generally confirm these detections.  Note that in many cases, two of
these features are often found superimposed atop a stellar \ion{He}{1}
line.\\

{\bf CH$^+$} -- About 10 of our sightlines also show absorption due to
CH$^+$. This species is detected at 3447.076\AA\ and/or 3579.020\AA.
These correspond to the $R$(0) line from the $4-1$ band, and the
$R$(0) line of the $3-0$ band \citep{1960APJ...131....1D}. While
CH$^+$ was originally thought to trace shocks, it is now thought to
trace turbulent dissipation regions \citep[see
  e.g.][]{2013RvMP...85.1021T}. \\

{\bf CN} -- CN is well known as an interstellar molecule for its
  potential to measure the temperature of the cosmic microwave
  background radiation by using the $B ^2\Sigma - X ^2\Sigma$ optical
  lines \citep{1985ApJ...297..119M}.  In our spectral range, three
  transitions in the (1,0) vibrational band -- the R(1), R(0), and
  P(1) lines -- are detected at $\lambda\lambda$ 3579.453, 3579.963,
  3580.937 towards a handful of targets.  Typically the P(1) line is
  not seen, likely because it is too weak -- it has the lowest
  oscillator strength of the three transitions
  \citep{1989ApJ...343L...1M}. \\

{\bf NH} -- NH was first detected in the near-UV by
\citet{1991ApJ...376L..49M} who detected both the R$_1$(0) and the
$^R$Q$_{21}$(0) transitions of the A$^3$$\Pi$-X$^3$$\Sigma$ system at
respectively 3358.0525\AA\ and 3353.9235\AA.  We detect these
transitions in $\sim$5 sightlines. \\

{\bf OH} -- The spectra of $\sim$10 targets also show absorption
due to OH. There are 5 $A-X$ lines associated with this species in our
wavelength range. However, we can only really observe three features:
a blend of the 3072.01\AA\ and 3072.0637\AA\ lines; a blend of the
3078.4399\AA\ and 3078.472\AA\ features; and the 3081.664\AA\ feature.
Of these, the 3072\AA~transitions R$_1$ (3/2) and R$_{Q_{21}}$ (3/2)
are the most recent to be discovered; all features are described by
\citet{1996ApJ...465L..57F}. The blends pose difficulties in providing
accurate measurements for the individual lines. \\

{\bf OH$^+$} -- OH$^+$ is one of the more recently detected
interstellar species \citep{2010A&A...518A..26W}. The near-UV line
near 3584\AA\ was previously described by various authors
\citep{2010ApJ...719L..20K,2011A&A...530A..26G,2014ApJ...781L...8P}. The
lack of velocity shifts between the neutral and ionized species rules
out a shock formation mechanism; \citet{2014ApJ...781L...8P} conclude
that cosmic ray ionization is the most likely route for this
species. We report the detections of the OH$^+$ feature in
BD~-14$^{\circ}$~5037.  \\

We could furthermore not detect the 3360\AA\ line of SH$^+$ in
our interstellar superspectrum, nor in any individual line of sight.

\subsection{Tentative Interstellar Features}

{\bf 3147.985\AA} -- When comparing the interstellar superspectrum of
reddened targets to its unreddened counterpart, a weak and relatively
broad feature is apparent near 3148\AA (see Fig.~\ref{Fig: RvsUR}).
This feature is not detected when comparing the reddened targets in
the various rest frames.  When measured in the interstellar
superspectrum, this feature has $W_{\lambda}$=1.0 $\pm$ 0.3 m\AA,
corresponding to a 3$\sigma$ detection. This may thus be a weak
interstellar feature. \\

\begin{figure}[t]
  \resizebox{\hsize}{!}{\includegraphics{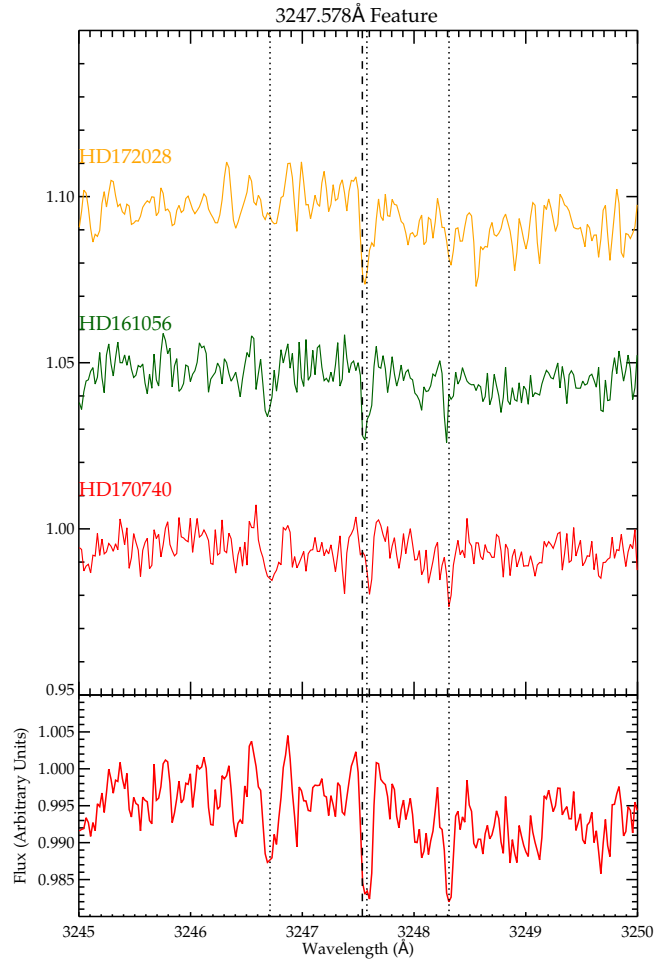}}
  \caption{\label{Fig:feature-3247}The optimal spectra of HD~172028,
    HD~161056 and HD~170740 in the interstellar rest frame near the
    rest wavelength of a \ion{Cu}{1} line at 3247.537\AA~(marked with a
    broken dashed line) show a feature at 3247.578\AA.
    In the same range, two more features are seen
    to line up in these targets at $\lambda\lambda$~3246.711, 3248.313,
    marked with a dotted line.  Each of these features is seen more
    clearly in the co-add of these three targets (bottom panel).}
\end{figure}

{\bf 3246.711, 3247.578, 3248.313 \AA} -- A weak feature aligns in the
spectra of  HD 161056, HD 170740 and HD 172028 at
3247.578\AA\ (see Fig.~\ref{Fig:feature-3247}). Coincidentally,
this is close to the wavelength of one of two ground state electronic
transitions of \ion{Cu}{1} in our wavelength range. If this line were
indeed a \ion{Cu}{1} line, we should also see the line at
3273.954\AA\ whose oscillator strength is  only a factor of 
  two lower \citep{1966ApJ...144..774M}. However, we do not see any
hint of an absorption line at that wavelength; thus, this line is not
likely to be the \ion{Cu}{1} line. On either side of this possible
interstellar lines are two even weaker lines that appear to line up at
3246.711\AA, 3248.313\AA. All these are marginal though -- they may or
may not be true interstellar lines. \\

\begin{figure}[t]
\resizebox{\hsize}{!}{\includegraphics{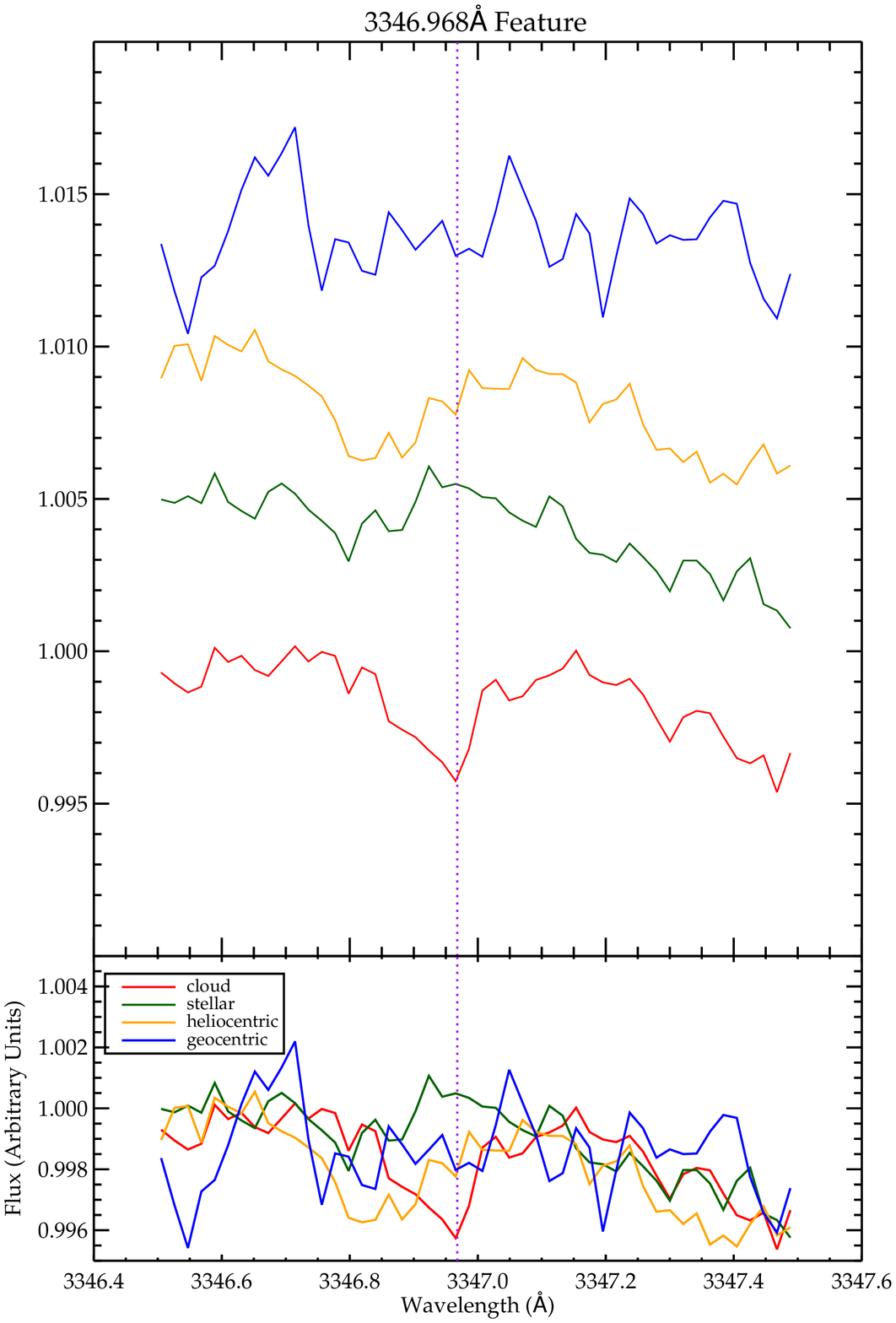}\includegraphics{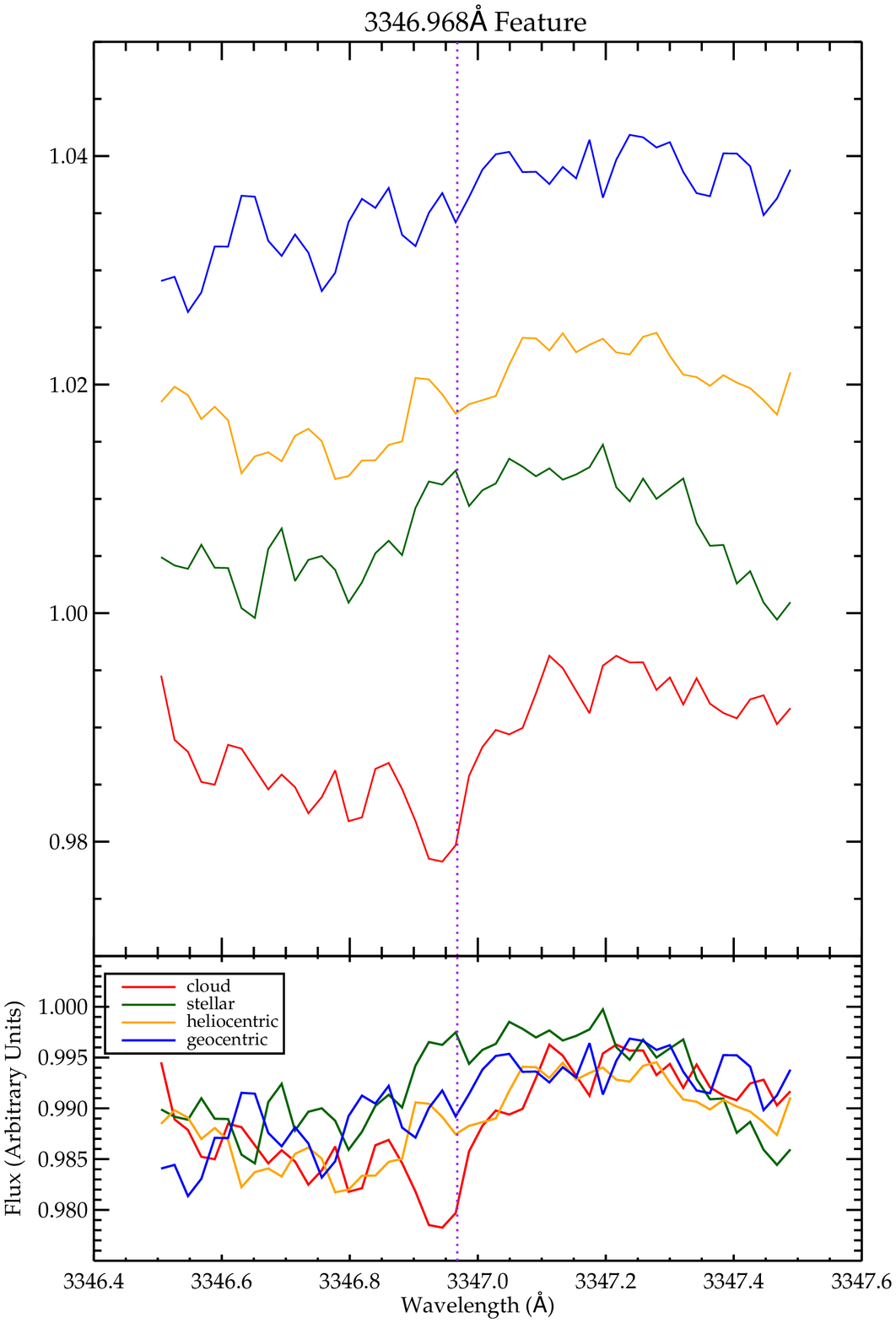}}
\caption{\label{Fig:feature-3346} The top panel shows the geocentric
  (blue), heliocentric (yellow), stellar (green) and interstellar (red)
  rest frame.  ({\em Left:}) A broad, asymmetric feature appears in
  the interstellar superspectrum (red) of all targets.  There is no
  corresponding atmospheric feature at this position. ({\em Right:})
  Focusing on the targets which clearly show this feature
  (i.e. HD~115363, HD~142758, HD~188220, HD~210121 and HD~143448, the
  asymmetric feature becomes much more regular in the interstellar
  frame. }
\end{figure}

{\bf 3346.968\AA} -- A weak, but conspicuous feature is apparent in
the interstellar superspectrum at 3346.968\AA\ (see
Fig.~\ref{Fig:feature-3346}). The feature is better defined still when
considering a limited sample of targets: HD~115363, HD~142758,
HD~188220, HD~210121, HD~143448 where we also tentatively see the
feature in the optimal target spectra. This feature also shows up
prominently in a comparison between reddened and unreddened
superspectra (see Fig.~\ref{Fig: RvsUR}). Therefore, this does appear
to be a true interstellar absorption feature. We have not been able to
find an identification for this feature. When measured in the
interstellar superspectrum, this feature has $W_{\lambda}$=0.5 $\pm$
0.1 m\AA, corresponding to a 5$\sigma$ detection.\\

{\bf 3362.942\AA} -- A weak and shallow feature may be present in the
interstellar superspectrum at 3362.942\AA. However, we do not detect
this feature in any individual object. Moreover, the geocentric
superspectrum exhibits two absorption features bracketing the possible
interstellar feature. Since interstellar velocities are fairly low, it
is possible that the apparent interstellar feature results from a
chance superposition of these telluric lines when shifting to the
interstellar rest frames.  On the other hand, the feature shows up
quite clearly when comparing the reddened to the unreddened
superspectrum (see Fig.~\ref{Fig: RvsUR}) thus strengthening the case
for an interstellar band. When measured in the interstellar
superspectrum, this feature has $W_{\lambda}$=0.6 $\pm$ 0.1 m\AA,
corresponding to a 6$\sigma$ detection. \\

\begin{figure}[t]
  \resizebox{\hsize}{!}{\includegraphics{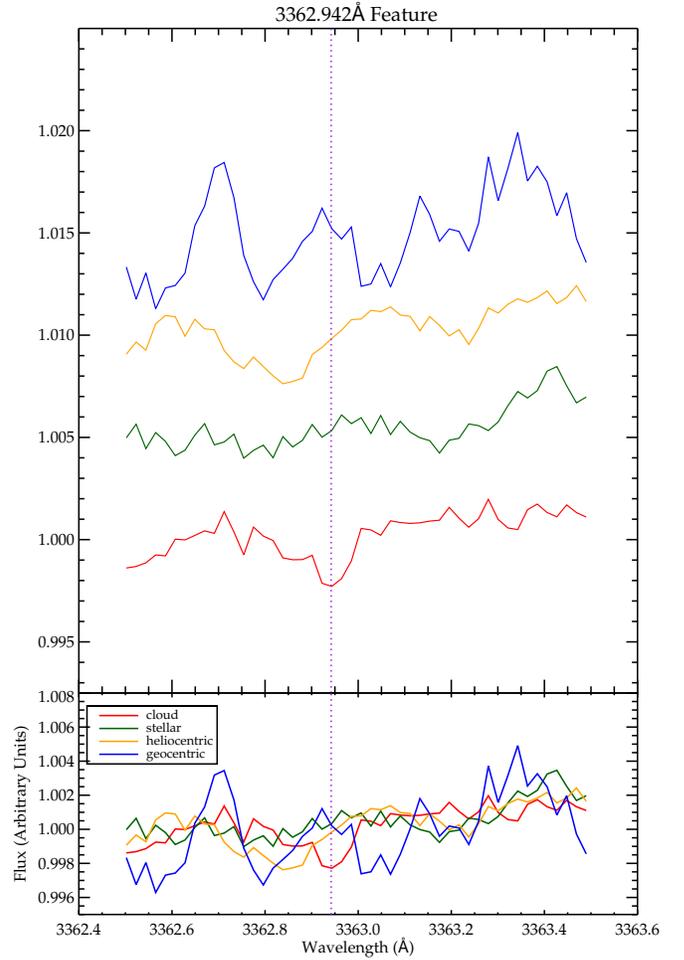}}
\caption{\label{Fig:feature-3362} The top panel shows the geocentric
  (top; blue), heliocentric (yellow), stellar (green) and interstellar
  (bottom; red) rest frame. A weak, diffuse band appears in the
  interstellar superspectrum (red; bottom spectrum of the top panel).
  Note the prominent atmospheric features are bracketing this possible
  interstelllar line. }
 \end{figure}

 {\bf 3572.656\AA} -- A slightly broader feature shows up in the
 interstellar superspectrum at 3572.656. We suspect that the strong
 telluric residuals in this range will have at least some contribution
 from this geocentric feature. However, the feature also shows up when
 comparing the reddened to the unreddened superspectrum (see
 Fig. \ref{Fig: RvsUR}). When measured in the interstellar
 superspectrum, its strength is comparable to the atomic, and
 molecular features which appear around 3578\AA (\ion{Cr}{1}, CN,
 CH$^{+}$): $W_{\lambda}$=1.1 $\pm$ 0.1 m\AA, corresponding to a 10$\sigma$
 detection. 

 \begin{figure}[t]
  \resizebox{\hsize}{!}{\includegraphics{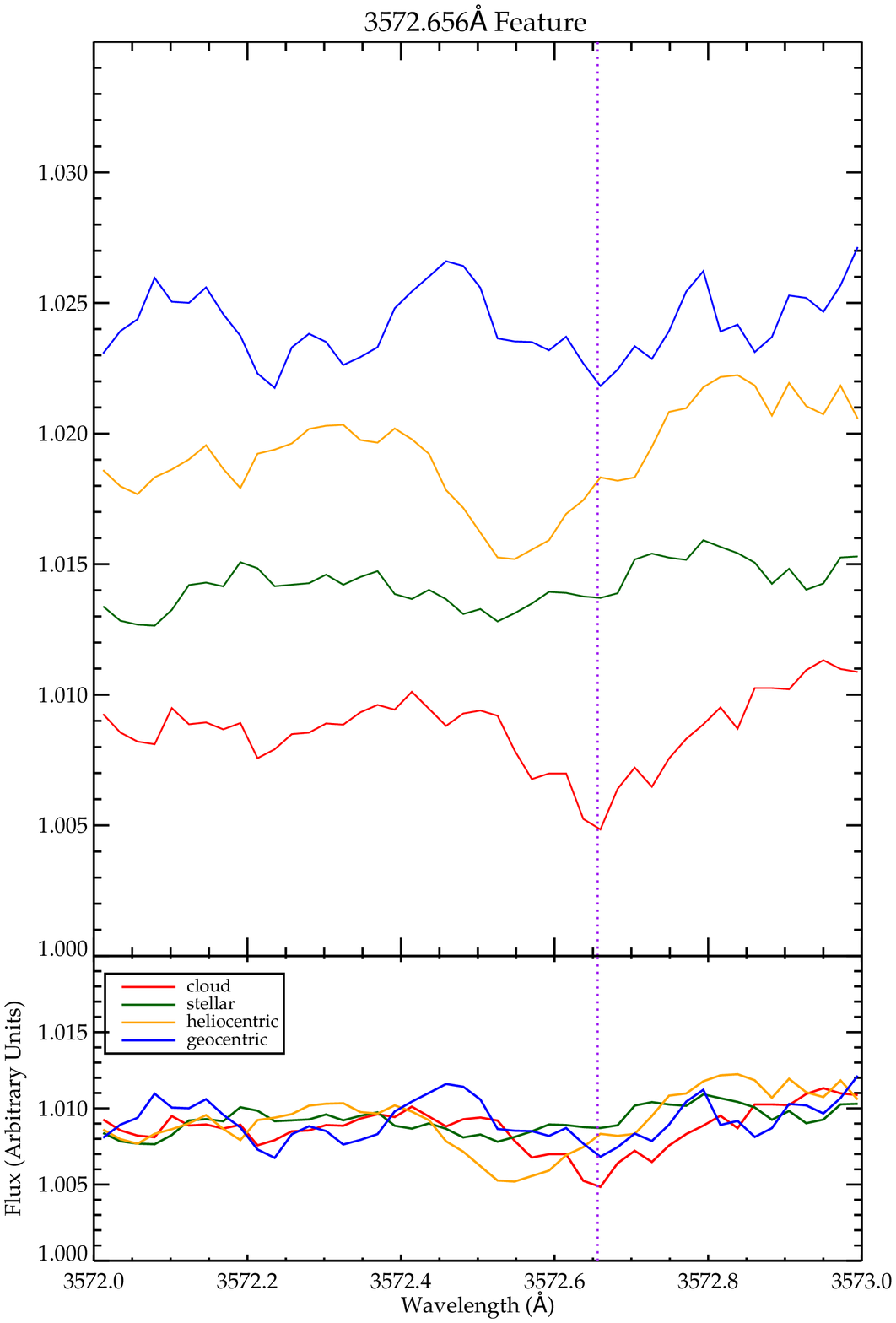}}
 \caption{\label{Fig:feature-3572} The top panel shows the geocentric
   (top; blue), heliocentric (yellow), stellar (green) and interstellar
   (bottom; red) rest frame.  A broad interstellar feature appears
   around 3572.656\AA.  It may have some geocentric contribution.}
\end{figure}

\subsection{Line of sight measurements}

We have measured basic line properties (central wavelengths,
  widths\footnote{Full Width at Half Maximum, FWHM} and equivalent
  widths) for each identified line that we could detect in the optimal
  spectrum of each of our targets. This was done by fitting (a)
  Gaussian profile(s) to the features in the interstellar rest
  frame. In many cases, this involved fitting several components
  originating from different clouds at the same time. We only aimed at
  reproducing the strongest components; thus, in some cases, weak
  residual shoulders are still visible in e.g. the \ion{Na}{1}
  lines. The \ion{Ti}{2} lines typically exhibit more components
  (often blended) over a wider velocity range than the \ion{Na}{1} or
  molecular lines; since many lines are also weak, this required a
  slightly different approach. For these lines, we first determined
  the number of cloud components by fitting the strongest line (at
  3383.768\AA). Once a satisfactory fit was obtained, we used the same
  components to fit the other lines while keeping the number of
  components as well as their central wavelengths and widths fixed
  (thus only allowing the strength of the lines to vary). Note that we
  do not account for any potential stellar absorption since this
  should not be a major contribution for early-type stars
  \citep{2010MNRAS.404.1321W}.

  The resulting line fits can be inspected in
  Fig.~\ref{Fig:measure_plot}; the measured values are listed in
  Table~\ref{tab:linesofsight}. Features that are not reported
  represent non-detections. The equivalent widths listed in
  Table~\ref{tab:linesofsight} are derived from the Gaussian fit
  parameters and uncertainties are the formal 1$\sigma$ uncertainties
  derived from the fit. For many of the \ion{Ti}{2} lines, we only
  list an uncertainty for the equivalent width since all other
  parameters were fixed as described above. For multi-component
  features, we also present the measured equivalent width obtained by
  integrating across the entire profile; these values are listed in
  square brackets in Table~\ref{tab:linesofsight}. The largest source
  of uncertainty on this measurement is the exact positioning of the
  local continuum. To get a good sense of this uncertainty and its
  effect on the measurement, we carried out a Monte Carlo experiment
  in which we reposition the continuum many times, commensurate with
  the local S/N, and integrate again. The average over all trials is
  what we used as the measured equivalent width in
  Table~\ref{tab:linesofsight}; the standard deviation over all runs
  is what is used as its uncertainty.

  We compared our equivalent widths of the CH $B-X$ system
  measurements to \citet{2011AN....332..167W}, and found that our
  values were generally in excellent agreement. We noted that
  \citet{2011AN....332..167W} reported some measurements for
  borderline detections (W$_{\lambda} < 3\sigma$) in sightlines where
  optical CH lines were quite strong.  For the sake of consistency, we
  have included these CH features as well. We also compared our
  results for the $\lambda\lambda$ 3072, 3241, 3383 \ion{Ti}{2}
  features to \citet{2010MNRAS.404.1321W} and again found excellent
  agreement. Note that these authors also include plots of \ion{Ti}{2}
  features in the heliocentric frame that are entirely consistent with
  our results. 

  It is interesting to note that only the \ion{Na}{1} doublet and the
  3241.994\AA\ and 3383.768\AA\ \ion{Ti}{2} features are observed in
  each target. On the other hand, only a handful of targets show
  \ion{Cr}{1}, NH, CH, or OH$^{+}$.  We also note that some targets
  show a feature coincident with the 3066.354\AA\ line of \ion{Ti}{2};
  however, this region of our spectra is very noisy. Similarly, we
  noticed systematic problems (artefacts) in many spectra around the
  CH feature at 3146.010\AA. This probably points to issues with the
  data reduction or CCD itself; therefore, measurements of this
  feature should be used with caution. 
  
  On first sight, one may expect to find that the presence (and
  strength) of these features should roughly increase with
  $E(B-V)$. This, however, is clearly not the case. For instance, the
  line of sight toward HD~183143 has $E(B-V)=1.21$ but shows only 12
  features, and many of those are weak. HD~154368 on the other hand
  has a much lower reddening ($E(B-V)=0.47$), but has many more
  features that are much stronger. While differences in abundances may
  play a role, such differences are primarily attributable to very
  different physical conditions that lead to different chemistry
  \citep[see e.g.][]{1996ApJ...465..245S}. This is reminiscent of the
  very clear differences in terms of molecular lines and DIBs observed
  in the lines of sight toward $\zeta$ Oph (HD 149757) and $\sigma$
  Sco (HD 147165) and attributable to differences in the physical
  conditions of the intervening clouds \citep[see
    e.g.][]{Cami:DIBcorrelations}.

%%%%% DISCUSSION %%%%%%%%%%%%%%%%%%%%%%%
\section{Discussion}
\label{Sec:discussion}

\begin{deluxetable}{llclrl}[t]
\tablecaption{\label{tab:PAHs}Upper Limits for the column densities of
  small neutral PAHs in the near-UV. }
\tablehead{%
\colhead{Species} &
\colhead{$\lambda$} &
\colhead{$f$} &
\colhead{$W_{\lambda, \rm max}$} &
\multicolumn{2}{c}{$N_{\rm max}$} \\
&
\colhead{[\AA]} &
\colhead{[10$^{-3}$]} &
[m\AA] &
\multicolumn{2}{c}{[$10^{12}$ cm$^{-2}$]}\\
& & & & (Here) & (Lit.)
}

\startdata
\hline
%From Gredel's paper
Anthracene         & 3610.74 & 16\tablenotemark{(a)}  & 0.47 &  0.25 & 0.53\tablenotemark{a}\\
Phenanthrene       & 3409.21 & .16\tablenotemark{(a)} & 0.40 & 24.90 & 62.0\tablenotemark{a}\\
Pyrene             & 3208.22 & 97\tablenotemark{(a)}  & 2.38 &  0.27 & 0.11\tablenotemark{a}\\
                   & 3166.26 & 33\tablenotemark{(a)}  & 2.75 &  0.92 & 0.34\tablenotemark{a}\\
2,3-benzofluorene  & 3344.16 & 8.5\tablenotemark{(a)} & 0.44 &  0.52 & 1.20\tablenotemark{a}\\
                   & 3266.63 & 2.5\tablenotemark{(a)} & 0.42 &  1.77 & 4.20\tablenotemark{a}\\
Benzo[ghi]perylene & 3512.15 & 9\tablenotemark{(a)}   & 2.18 &  2.20 & 1.00\tablenotemark{a}\\
                   & 3501.76 & 5\tablenotemark{(a)}   & 1.75 &  3.08 & 1.80\tablenotemark{a}\\ \hline
%From Salama's paper
Acenaphthene        & 3175    & 2.2\tablenotemark{b}  & 0.78 &	3.99 & 3.00\tablenotemark{b}\\
Benzo[ghi]perylene  & 3689.4  & 150\tablenotemark{c}  & 0.71 &	0.04 & 0.12\tablenotemark{b}\\
2-methylnaphthalene & 3152.7  & 2.3\tablenotemark{b}  & 0.80 &	3.93 & 4.90\tablenotemark{b}\\
Phenanthrene        & 3408.4  & 23\tablenotemark{b}   & 0.66 &	0.28 & 0.21\tablenotemark{b}\\
Pyrene              & 3205.8  & 160\tablenotemark{b}  & 2.56 &	0.18 & 0.97\tablenotemark{b}\\
Benzofluorene       & 3345.1  & 24\tablenotemark{d}   & 0.29 &	0.12 & 0.69\tablenotemark{b}\\
Anthracene          & 3611.8  & 10\tablenotemark{e}   & 0.32 &	0.28 & 2.60\tablenotemark{b}\\ \hline
%From Tim's paper
Triphenylene	  &3293.52 & 0.7\tablenotemark{f} &	0.43 &	6.56 &
\tablerefs{
(a) \citet{2011A&A...530A..26G}; (b) \citet{2011ApJ...728..154S};
(c) \citet{2005JChPh.123a4312T}; (d) \citet{2006JMoSt.786..105S}; (e) \citet{HermineThesis};(f) \citet{2007JChPh.126h4304K}}
\enddata
\end{deluxetable}

\begin{figure}[t]
\resizebox{\hsize}{!}{\includegraphics{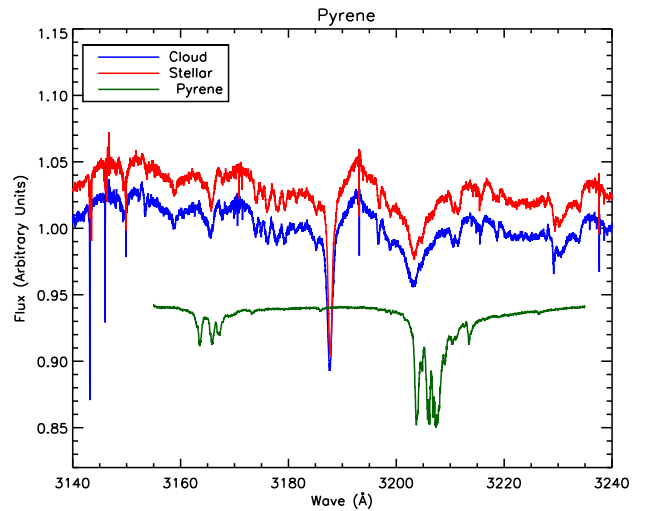}}
\caption{\label{Fig:pyrene} A laboratory spectrum of pyrene (bottom;
  green) is compared to both the stellar (top; red), and interstellar
  superspectra (middle; blue).  Around 3160\AA, all spectra have a
  series of features. However, there is no similarity between the
  features in pyrene and the others.  }
\end{figure}

\begin{figure*}[t]
  \begin{tabular}{c}
  \resizebox{\hsize}{!}{\includegraphics{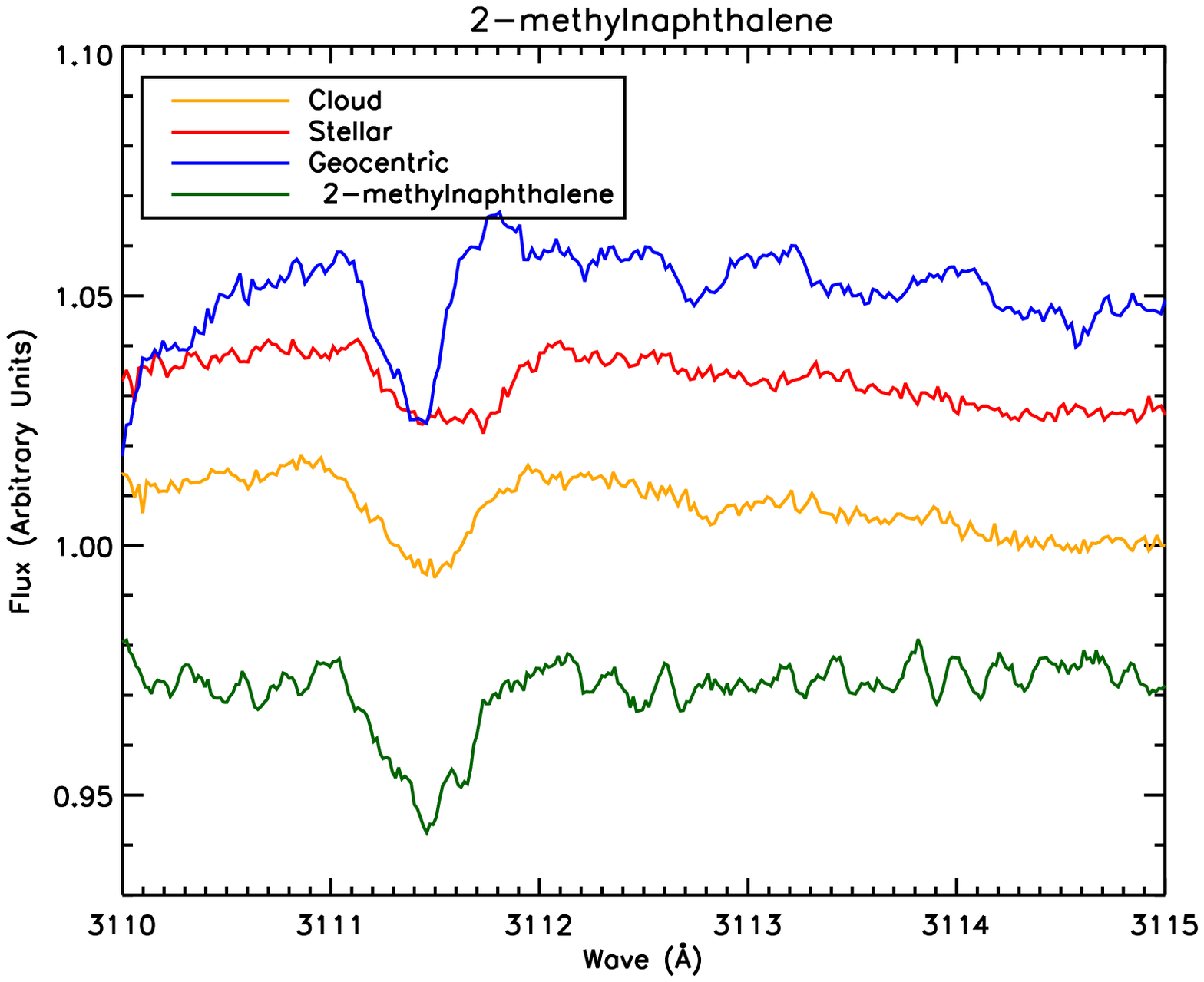}
    \includegraphics{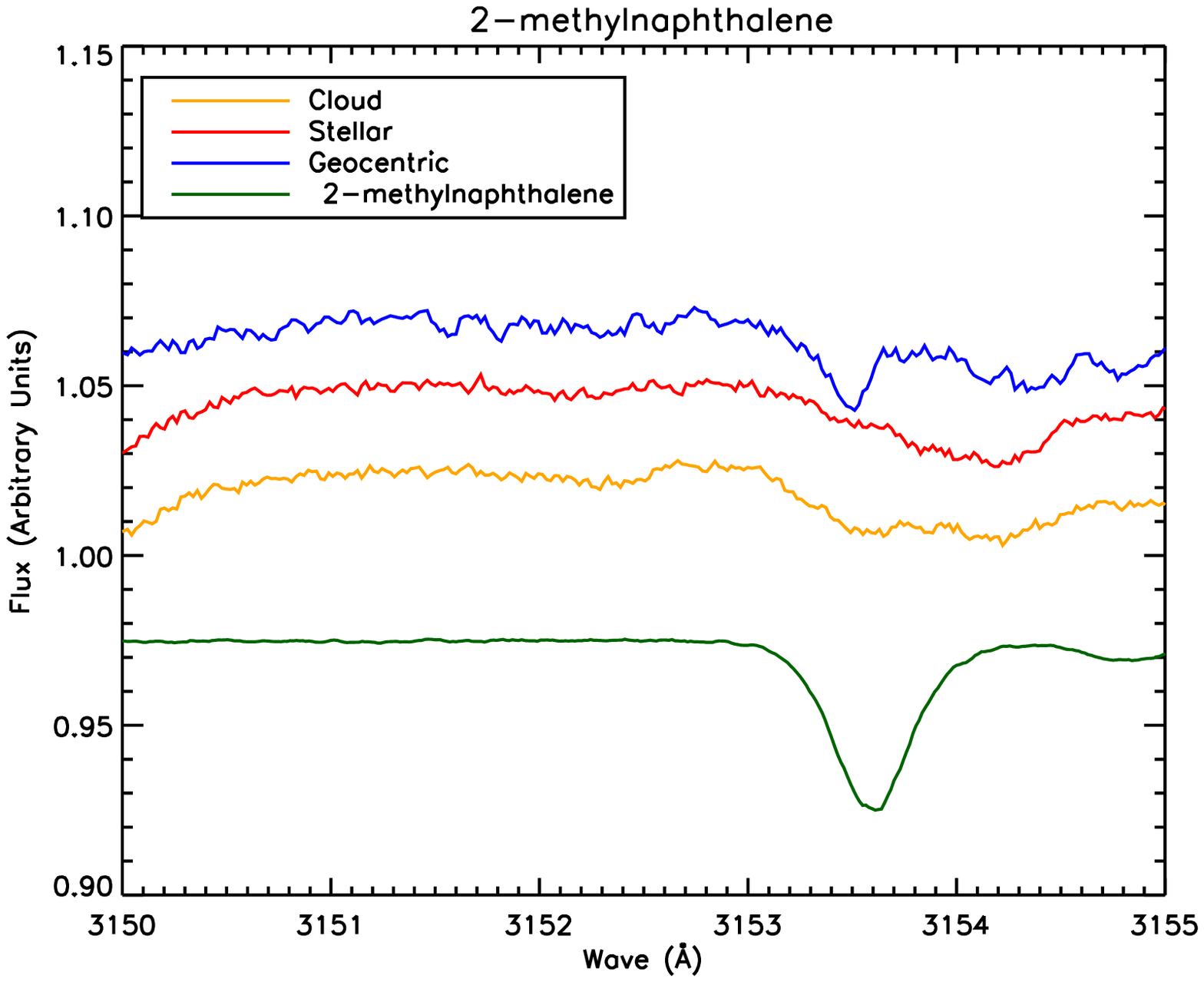}} \\

  \resizebox{\hsize}{!}{\includegraphics{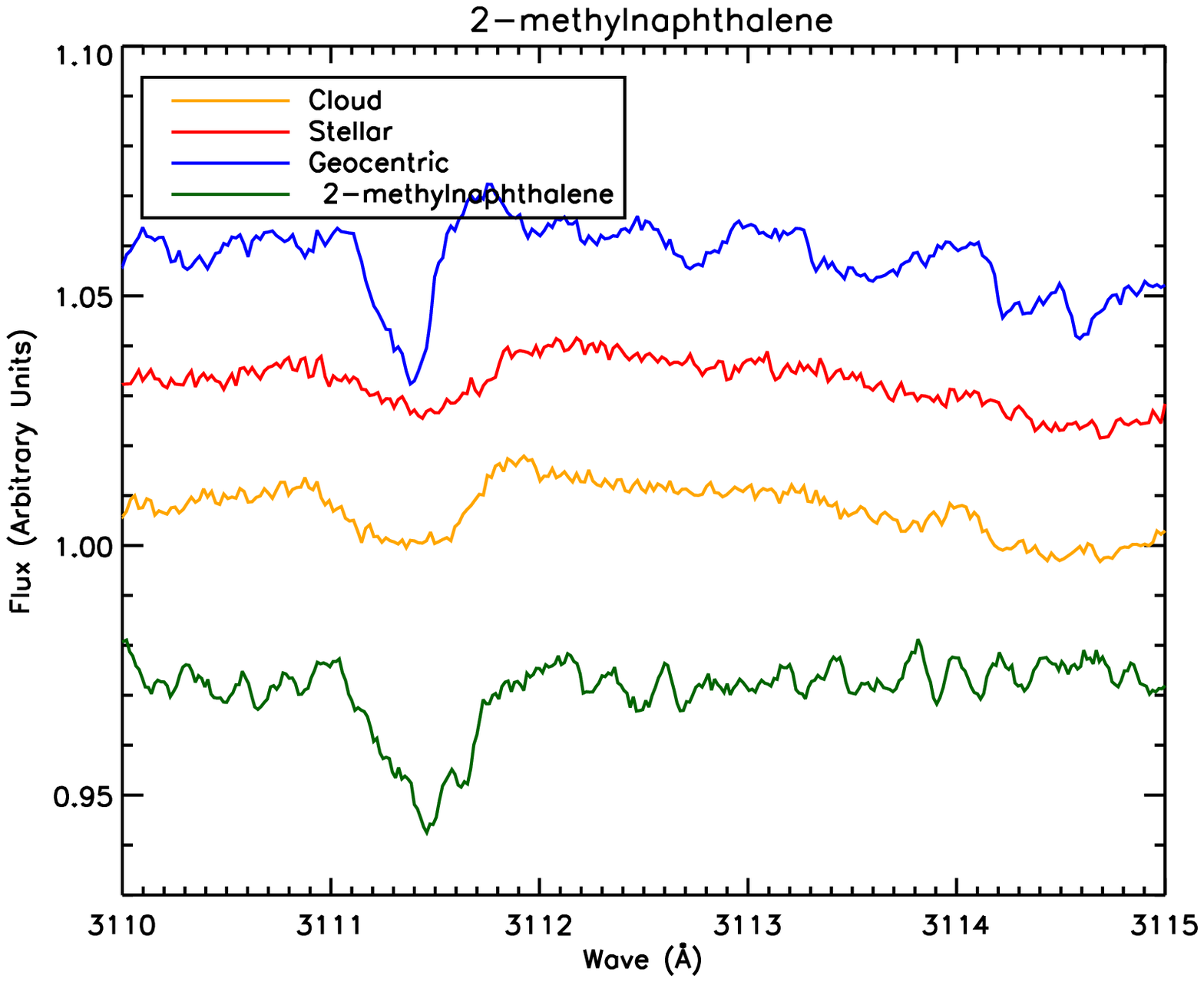}
    \includegraphics{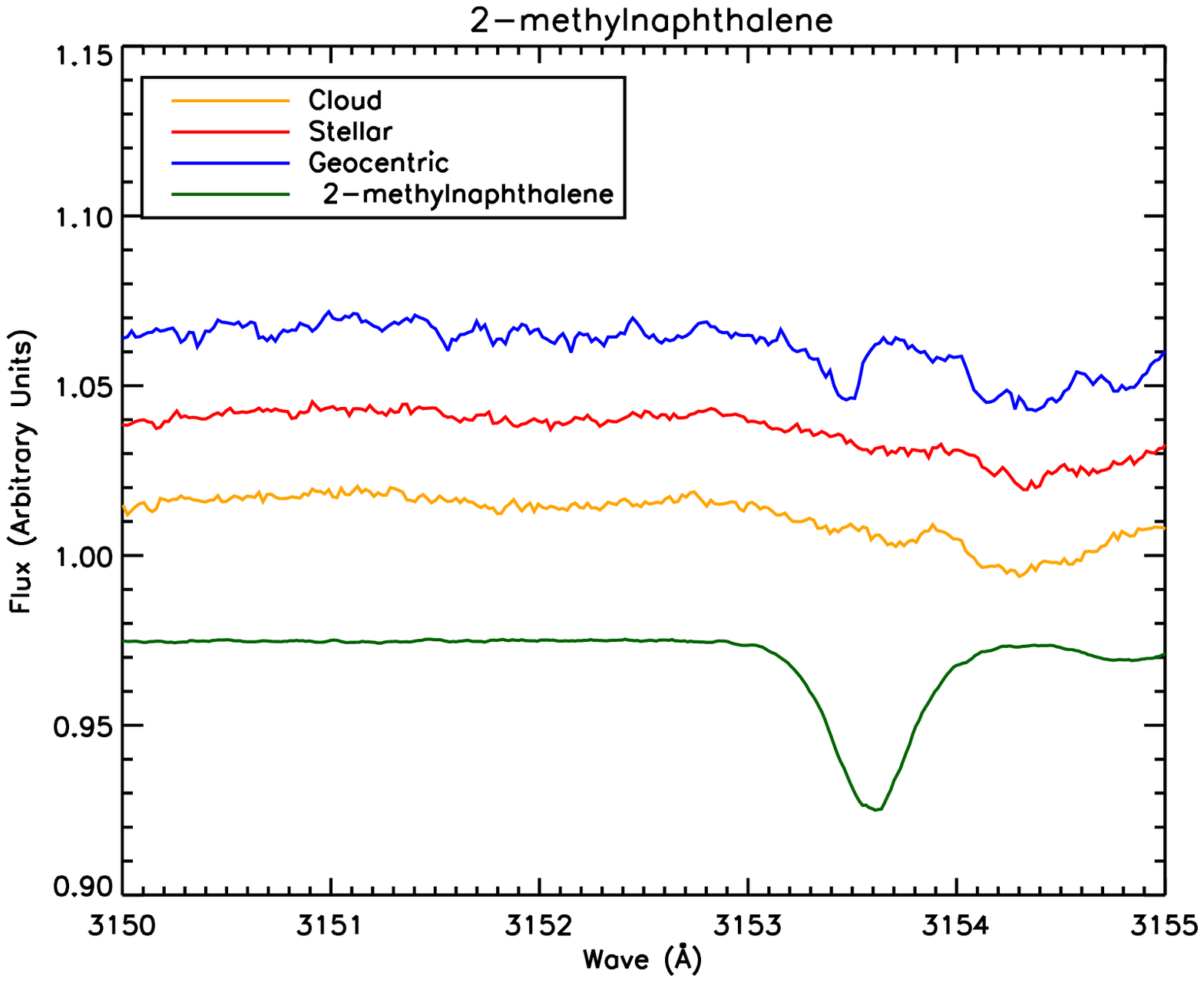}} \\
  \end{tabular}
  \caption{\label{Fig:natph} A laboratory spectrum of
    2-methylnatphalene in the region of two of its strongest
    transitions near 3111.5\AA\ (left panels) and 3152.7\AA\ (right
    panels) is compared to the superspectra of the reddened targets
    (top panels), and unreddened targets (bottom panels). Each figure
    shows the geocentric (top; blue), stellar (red), and interstellar
    (orange) superspectra compared to the (scaled) laboratory spectrum
    (bottom; green). At both wavelengths, a feature appears in the
    interstellar spectrum with approximately the same characteristics
    (wavelength, width) as the laboratory spectrum. However, this
    feature is much better defined (narrower) in the geocentric
    superspectrum, suggesting that this feature is telluric in nature
    -- possibly 2-methylnaphtalene in the earth's atmosphere. }
\end{figure*}

It is reassuring that we detect tens of known interstellar atomic and
molecular lines in this wavelength range, even if they do not appear
in the majority of the individual targets. For instance, while we
detect \ion{Na}{1} in all 51 sightlines, the \ion{Cr}{1} lines are
only found in a handful ($<10$) of sightlines, yet they do show up in
our interstellar superspectrum. This confirms that our method is
indeed sensitive to most narrow interstellar lines. We are thus
confident that if a (narrow) feature was somewhat common in the ISM,
we should have detected it. However, our method does not work well for
broad features, and thus we may not be sensitive to see broad,
DIB-like absorption features. This is an important caveat to bear in
mind for the discussion that follows.

Apart from known interstellar species, our survey results in only 7
possible new features that could be atomic or molecular lines, or
narrow DIBs. If the latter is the case, such a low number is in line
with the decreasing density of DIBs toward the blue at optical
wavelengths \citep[see e.g.][and references
  therein]{2014IAUS..297..173W}. Thus, there are at most very few
unknown interstellar lines left in the near-UV range.

The absence of features in a sensitive survey can be used to determine
upper limits to the column densities of molecular species whose
gas-phase spectra are known, and that could possibly be present in the
interstellar medium. Following the approach of
\citet{2011ApJ...728..154S} and \citet{2011A&A...530A..26G}, we
determined upper limits to the average ISM column densities of several
PAHs from the aforementioned works, as well as from
\citet{2007JChPh.126h4304K}; the results are listed in
Table~\ref{tab:PAHs}. Note that this is the first time that
triphenylene has been compared to observations. These upper limits are
of the same order of magnitude as those listed by the previous
authors. Note though that the peak positions and oscillator
strengths of phenanthrene, pyrene, benzofluorene and anthracene used
by \citet{2011ApJ...728..154S} and \citet{2011A&A...530A..26G} differ
noticeably from one another. 

Since we have used many more targets than both
\citet{2011ApJ...728..154S} and \citet{2011A&A...530A..26G}, we would
expect our upper limits to be more
significant. \citet{2011A&A...530A..26G} use a detection limit of
1m\AA~for all bands.  This strategy is a reasonable to
first-order. However, we find that the S/N decreases towards the blue
(see Figures \ref{Fig:superspec}).  Therefore, a more realistic upper
limit is obtained by measuring the S/N in the region of the feature --
this is the method we followed. \citet{2011ApJ...728..154S} also
followed this procedure.  However, we noticed that their quoted S/N of
their co-adds are comparable, and in some cases, significantly
higher than those from our interstellar superspectrum, notwithstanding
the fact that we use many more reddened targets than their 17 lines of
sight.

For comparison, we have created a co-add that includes 12 of their 17
targets (the other 5 are not in our target list) and show a part of
the spectrum in Fig.~\ref{Fig:pyrene}; this should be compared with
their Fig.~11. It is clear that the two are rather different; where
our co-adds show many stellar lines, there is no trace of stellar
lines in the Figures by \citet{2011ApJ...728..154S}, and it is not
clear how the authors removed these. We attempted to fit the stellar
lines ourselves, and inevitably found that this procedure resulted in
clear residuals near the line cores; these residuals did bear some
resemblance to the pyrene laboratory spectra. From our superspectra,
we find no evidence for interstellar lines in this range though. Note
also that \citet{2011ApJ...728..154S} state that their stellar spectra
are free of telluric features, which would otherwise interfere with
detecting weak features; as we have shown, there are many weak
telluric residuals left in our spectra. 

Finally,  we point out an intriguing case which exemplifies the
need to carefully consider features in the geocentric frame. When
comparing our superspectra to the laboratory spectrum of
2-methylnaphthalene, we noticed that there was a fairly good
correspondence between the astronomical and the laboratory spectra
(see Fig.~\ref{Fig:natph}). Indeed, both near 3111.5\AA\ and
3153.5\AA, we found absorption bands at about the location of the
strongest transitions of 2-methylnaphthalene. However, when
considering the geocentric superspectrum, those bands appear much
narrower and much more defined. This then suggests that the feature is
telluric in origin -- possibly, we are seeing 2-methylnaphthalene in
the earth's atmosphere. It is worth noting that small PAHs,
including alkylated naphthalenes are detectable in ambient air
\citep{1989EnST...23..723Z}.

While \citet{2011A&A...530A..26G} and \citet{2011ApJ...728..154S}
obtained upper limits for specific wavelengths and for a few neutral
PAHs, our survey shows that similar upper limits hold for the entire
wavelength range studied here (with of course the exception of the 7
unidentified features).  Thus, there is not much room for molecular
species that have narrow absorption bands. Only very broad (tens of
\AA) and shallow bands could possibly still be present in this part of
the spectrum.

\section{Conclusions}
\label{Sec:conclusion}

We have carried out a sensitive survey for interstellar features in
the near-UV (from 3050--3700\AA) using archival UVES data and a method
that compares various co-adds corresponding to different rest
frames. Our survey has revealed 30 known interstellar atomic and
molecular features, and 7 weak features that appear interstellar in
nature, but are unidentified. The high sensitivity of our survey
allows us to determine significant upper limits to the column
densities of various molecular carriers, and over the entire
wavelength range studied. Thus, there are only very few narrow and
unidentified interstellar features in the 3050--3700 \AA\ range.

\acknowledgments We would like to thank Steve Federman and Dan Welty
for assisting us with the identification of some transitions. We are
grateful to the referee for the constructive comments that have
improved this paper. We acknowledge support from the Natural Sciences
and Engineering Research Council of Canada (NSERC) through a Discovery
grant and an Undergraduate Student Research Award (USRA). \\

\clearpage

\appendix

%%%%%%%%%%%%%%%%%%%%%%%%%%%%%%%%%%%%%%%%%%%%%%%%%
%%%%%%%%%%%%%%%%%%%%%%%%%%%%%%%%%%%%%%%%%%%%%%%%%
%%%%%%%%%%%%%%%%%%%%%%%%%%%%%%%%%%%%%%%%%%%%%%%%%
%   Targets Table
%%%%%%%%%%%%%%%%%%%%%%%%%%%%%%%%%%%%%%%%%%%%%%%%%
%%%%%%%%%%%%%%%%%%%%%%%%%%%%%%%%%%%%%%%%%%%%%%%%%
%%%%%%%%%%%%%%%%%%%%%%%%%%%%%%%%%%%%%%%%%%%%%%%%%

\LongTables
\begin{deluxetable*}{lcrrr@{.}lrr@{.}lrcrc}[floatfix]%p{5cm}c}
\tablecaption{\label{Table:astrodata}Basic data of our targets and   observations. 
  \\}

%\tablewidth{0pt} 

\tablehead{\colhead{Object} 
  & \colhead{Sp.Type\tablenotemark{1}} &
  \multicolumn{1}{c}{RA} & 
  \multicolumn{1}{c}{DEC} &
  \multicolumn{2}{c}{$E_{B-V}$} 
  & \colhead{$v_{\star}$\tablenotemark{1}}
  & \multicolumn{2}{c}{$v_{\rm ISM}$\tablenotemark{2}}
  &\colhead{\# Obs\tablenotemark{3}} 
  & \colhead{Prog.~IDs} 
  &\colhead{S/N\tablenotemark{4}} 
  & \colhead{Type\tablenotemark{5}} \\ 
& 
& \multicolumn{1}{c}{[J2000]} 
&  \multicolumn{1}{c}{[J2000]}
& \multicolumn{2}{c}{[mag]} 
&  [km$\cdot$s$^{-1}$] 
& \multicolumn{2}{c}{[km$\cdot$s$^{-1}$]} 
&
&
&
&
}
 \startdata 
 HD 36861 & O8III  & 05 35 08.27 & +09 56 02.96 &0&09\tablenotemark{i} &33&25&2~(26.8\tablenotemark{aa} )&6& 266.D-5655  &141&R\\
HD 67536 & B2.5V  & 08 04 42.93 & -62 50 10.84 &0&12\tablenotemark{c} &0&14&3&3& 266.D-5655  &337&R\\
HD 68761 & B0.5III  & 08 12 27.36 & -36 59 21.23 &0&13\tablenotemark{c} &43&18&3&2& 266.D-5655  &245&R\\
HD 94963 & O6.5III  & 10 56 35.79 & -61 42 32.22 &0&24\tablenotemark{h}  &-3&10&1&2& 266.D-5655  &159&R\\
HD 96917 & O8.5Ib  & 11 08 42.61 & -57 03 56.91 &0&32\tablenotemark{i} &9&-15&8&2& 266.D-5655  &231&R\\
HD 105056 & O9.5Ia  & 12 05 49.87 & -69 34 22.98 &0&36\tablenotemark{b}  &1&17&5&2& 266.D-5655  &188&R\\
HD 105071 & B6Ia/Ib  & 12 05 53.61 & -65 32 48.75 &0&19\tablenotemark{c}  &-12&0&5&2& 266.D-5655  &136&R\\
HD 106068 & B8Ia/Ib  & 12 12 21.98 & -62 57 02.78 &0&26\tablenotemark{c}  &-2&7&3&2& 266.D-5655  &111&R\\
\enddata
\end{deluxetable*}

%%%%%%%%%%%%%%%%%%%%%%%%%%%%%%%%%%%%%%%%%%%%%%%%%
%%%%%%%%%%%%%%%%%%%%%%%%%%%%%%%%%%%%%%%%%%%%%%%%%
%%%%%%%%%%%%%%%%%%%%%%%%%%%%%%%%%%%%%%%%%%%%%%%%%
%   LINE OF SIGHT Measurements (TABLE)
%%%%%%%%%%%%%%%%%%%%%%%%%%%%%%%%%%%%%%%%%%%%%%%%%
%%%%%%%%%%%%%%%%%%%%%%%%%%%%%%%%%%%%%%%%%%%%%%%%%
%%%%%%%%%%%%%%%%%%%%%%%%%%%%%%%%%%%%%%%%%%%%%%%%%
\clearpage

\begin{deluxetable}{%
l% Target
l% Species    
l% lambda +/- error
r%@{}|% v_helio +/-
%p{3px}|%pm
@{~}l% error on v_helio
r@{.}l@{~$\pm$}%Equivalent Width +/-
r@{.}l% Error W
r@{.}l@{~}% % FWHM (mA) +/-
r%Error on FWHM
r@{.}l@{~}% FWHM km/s +/-
r% Error on FWHM
} 

\tablecaption{\label{tab:linesofsight}Lines Of Sight Measurements}

\tablehead{
 \multicolumn{1}{c}{\centering Target}      &
 \multicolumn{1}{c}{\centering Species}     &
 \multicolumn{1}{c}{${\rm \lambda_{helio}}$}  &
 \multicolumn{2}{c}{${\rm v_{helio}}$}        &
 \multicolumn{4}{c}{${\rm W_{\lambda}}$}       &  
 \multicolumn{6}{c}{FWHM} \\
 ~                                                                         & 
 ~                                                                        &
 \multicolumn{1}{c}{[\AA]}             &
 \multicolumn{2}{c}{[km s$^{-1}$]}        &
 \multicolumn{4}{c}{[m\AA]}            &
 \multicolumn{3}{c}{[m\AA]} &
 \multicolumn{3}{c}{[km~s$^{-1}$]}
}

\startdata
BD-14$^{\circ}$ 5037 & \ion{Cr}{1} & 3578.720~$\pm$~0.004 & 3.1~$\pm$ & 0.3 & 1&39 & 0&07 & 115&0~$\pm$ & 8.4 & 9&64~$\pm$ & 0.70 \\ 
         & \ion{Fe}{1} & 3440.522~$\pm$~0.004 & -7.3~$\pm$ & 0.4 & 1&14 & 0&07 & 60&0~$\pm$ & 10.5 & 5&23~$\pm$ & 0.92 \\ 
         &  ~  & 3440.665~$\pm$~0.004 & 5.2~$\pm$ & 0.4 & 4&03 & 0&28 & 139&1~$\pm$ & 11.7 & 12&12~$\pm$ & 1.02 \\ 
~ & ~ & \multicolumn{1}{c}{ ~ } &   &     & [5&11 & 0&75] & \multicolumn{1}{c}{~} & \multicolumn{5}{c}{~ } \\ 
         & \ion{Na}{1} & 3302.289~$\pm$~0.002 & -7.2~$\pm$ & 0.2 & 46&86 & 0&22 & 74&3~$\pm$ & 0.6 & 6&75~$\pm$ & 0.06 \\ 
         &  ~  & 3302.424~$\pm$~0.002 & 5.1~$\pm$ & 0.2 & 106&05 & 0&36 & 120&5~$\pm$ & 0.7 & 10&94~$\pm$ & 0.06 \\ 
~ & ~ & \multicolumn{1}{c}{ ~ } &   &     & [156&68 & 1&08] & \multicolumn{1}{c}{~} & \multicolumn{5}{c}{~ } \\ 
         & \ion{Na}{1} & 3302.904~$\pm$~0.002 & -6.7~$\pm$ & 0.2 & 33&94 & 0&14 & 61&5~$\pm$ & 0.6 & 5&58~$\pm$ & 0.05 \\
\end{deluxetable}

\clearpage
%%%%%%%%%%%%%%%%%%%%%%%%%%%%%%%%%%%%%%%%%%%%%%%%%
%%%%%%%%%%%%%%%%%%%%%%%%%%%%%%%%%%%%%%%%%%%%%%%%%
%%%%%%%%%%%%%%%%%%%%%%%%%%%%%%%%%%%%%%%%%%%%%%%%%
%   SuperSpectra Plot over full wavelength range
%%%%%%%%%%%%%%%%%%%%%%%%%%%%%%%%%%%%%%%%%%%%%%%%%
%%%%%%%%%%%%%%%%%%%%%%%%%%%%%%%%%%%%%%%%%%%%%%%%%
%%%%%%%%%%%%%%%%%%%%%%%%%%%%%%%%%%%%%%%%%%%%%%%%%
\figsetstart
\figsetnum{14}
\figsettitle{Comparison of geocentric, heliocentric, interstellar, and stellar rest frame superspectra}

\figsetend

\begin{figure}[h]
\figurenum{14}
\plotone{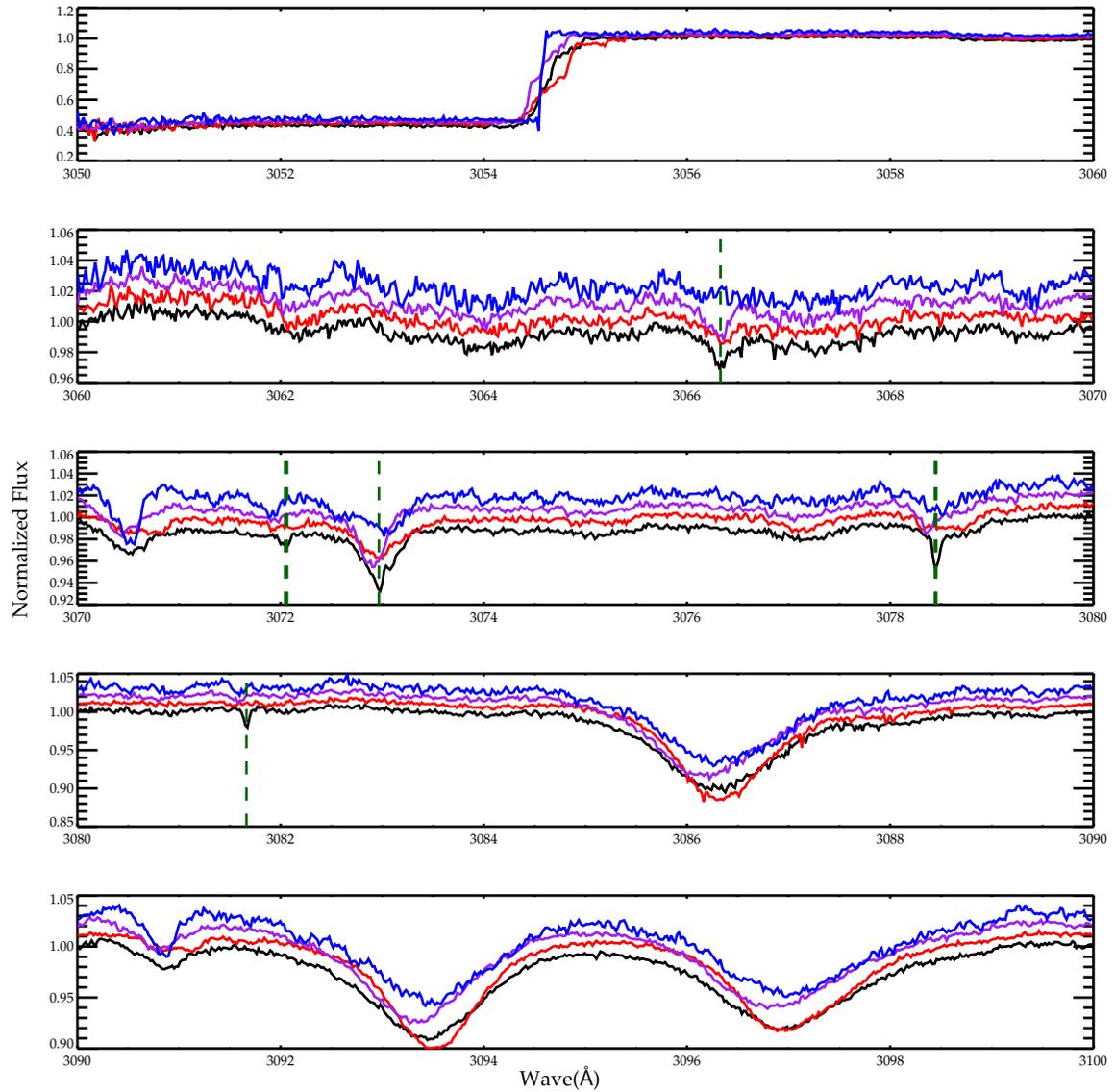}
\caption{\label{Fig:superspec}From top to bottom, the reddened superspectra (comprised of all 51 targets) are compared:
  geocentric (top; blue); heliocentric (purple); stellar (red); interstellar (bottom; black).
  Known interstellar lines (if any) are indicated with green vertical line, and
  features which appear interstellar in nature (if any) are shown with a red 
  vertical line.
  { Figures 14.1--14.13 are available in the online version of the Journal.}}
\end{figure}

\clearpage
%%%%%%%%%%%%%%%%%%%%%%%%%%%%%%%%%%%%%%%%%%%%%%%%%
%%%%%%%%%%%%%%%%%%%%%%%%%%%%%%%%%%%%%%%%%%%%%%%%%
%%%%%%%%%%%%%%%%%%%%%%%%%%%%%%%%%%%%%%%%%%%%%%%%%
%   Reddened vs. Unreddened standards
%%%%%%%%%%%%%%%%%%%%%%%%%%%%%%%%%%%%%%%%%%%%%%%%%
%%%%%%%%%%%%%%%%%%%%%%%%%%%%%%%%%%%%%%%%%%%%%%%%%
%%%%%%%%%%%%%%%%%%%%%%%%%%%%%%%%%%%%%%%%%%%%%%%%%
\figsetstart
\figsetnum{15}
\figsettitle{Comparison of unreddened and reddened interstellar rest frame superspectra}

\figsetend

\begin{figure}[h]
  \figurenum{15}
  \plotone{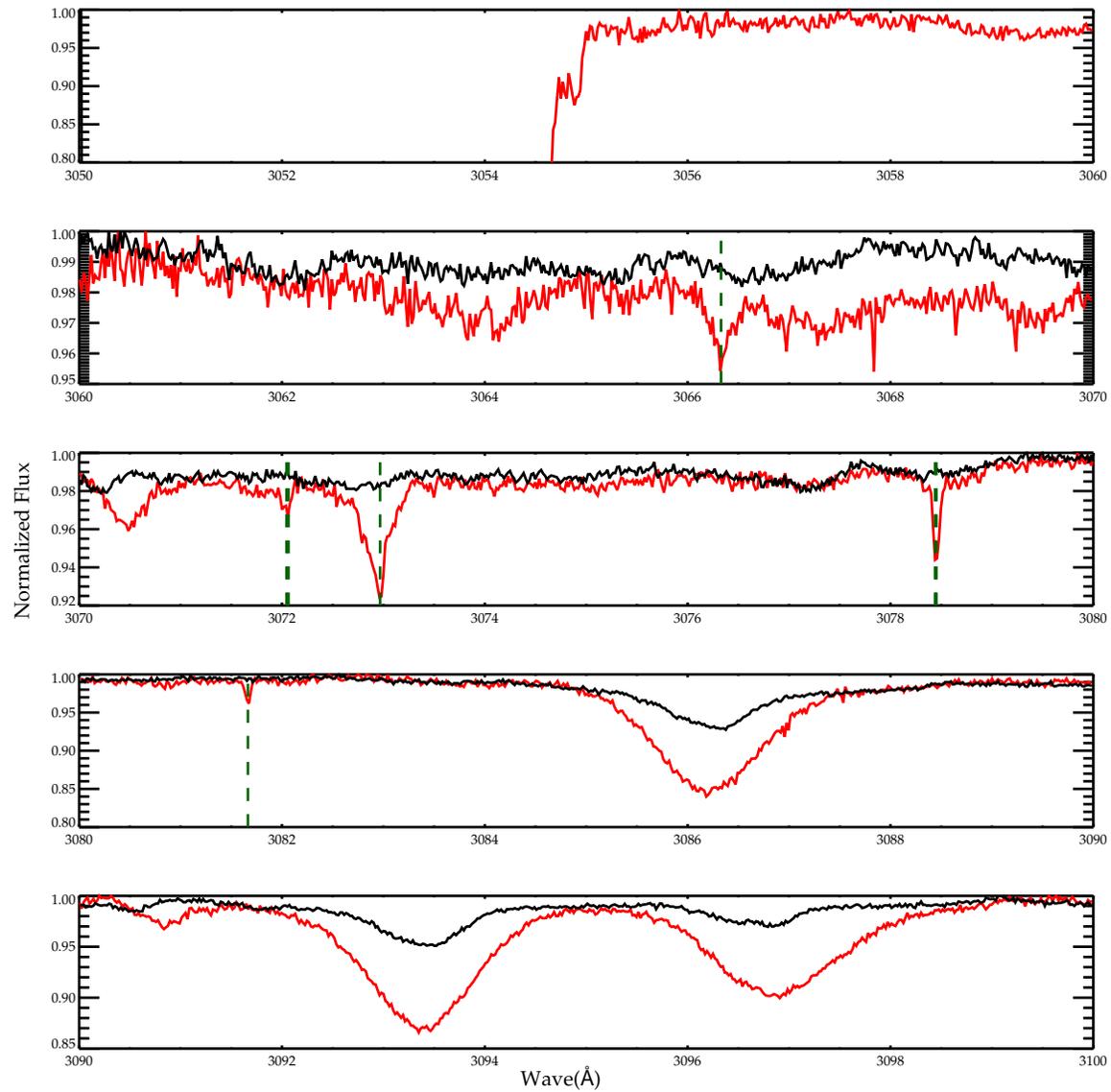}
  \caption{ \label{Fig: RvsUR} A co-add of reddened targets (red) with good spectral
    standards, are compared to a co-add of their spectral standards (black) in steps of 10\AA.
    Known interstellar lines (if any) are indicated with green vertical line, and 
    features which appear interstellar in nature (if any) are shown with a red vertical line.
    { Figures 15.1--15.13 are available in the online version of the Journal.  }}
\end{figure}

 \clearpage
%%%%%%%%%%%%%%%%%%%%%%%%%%%%%%%%%%%%%%%%%%%%%%%%%
%%%%%%%%%%%%%%%%%%%%%%%%%%%%%%%%%%%%%%%%%%%%%%%%%
%%%%%%%%%%%%%%%%%%%%%%%%%%%%%%%%%%%%%%%%%%%%%%%%%
%   LINE OF SIGHT Measurements (PLOTS)
%%%%%%%%%%%%%%%%%%%%%%%%%%%%%%%%%%%%%%%%%%%%%%%%%
%%%%%%%%%%%%%%%%%%%%%%%%%%%%%%%%%%%%%%%%%%%%%%%%%
%%%%%%%%%%%%%%%%%%%%%%%%%%%%%%%%%%%%%%%%%%%%%%%%%

%%%% Line of Sight Measurements....
\figsetstart
\figsetnum{16}
\figsettitle{Line of Sight Measurements}

\figsetend

\begin{figure}[h]
\figurenum{16}
\plotone{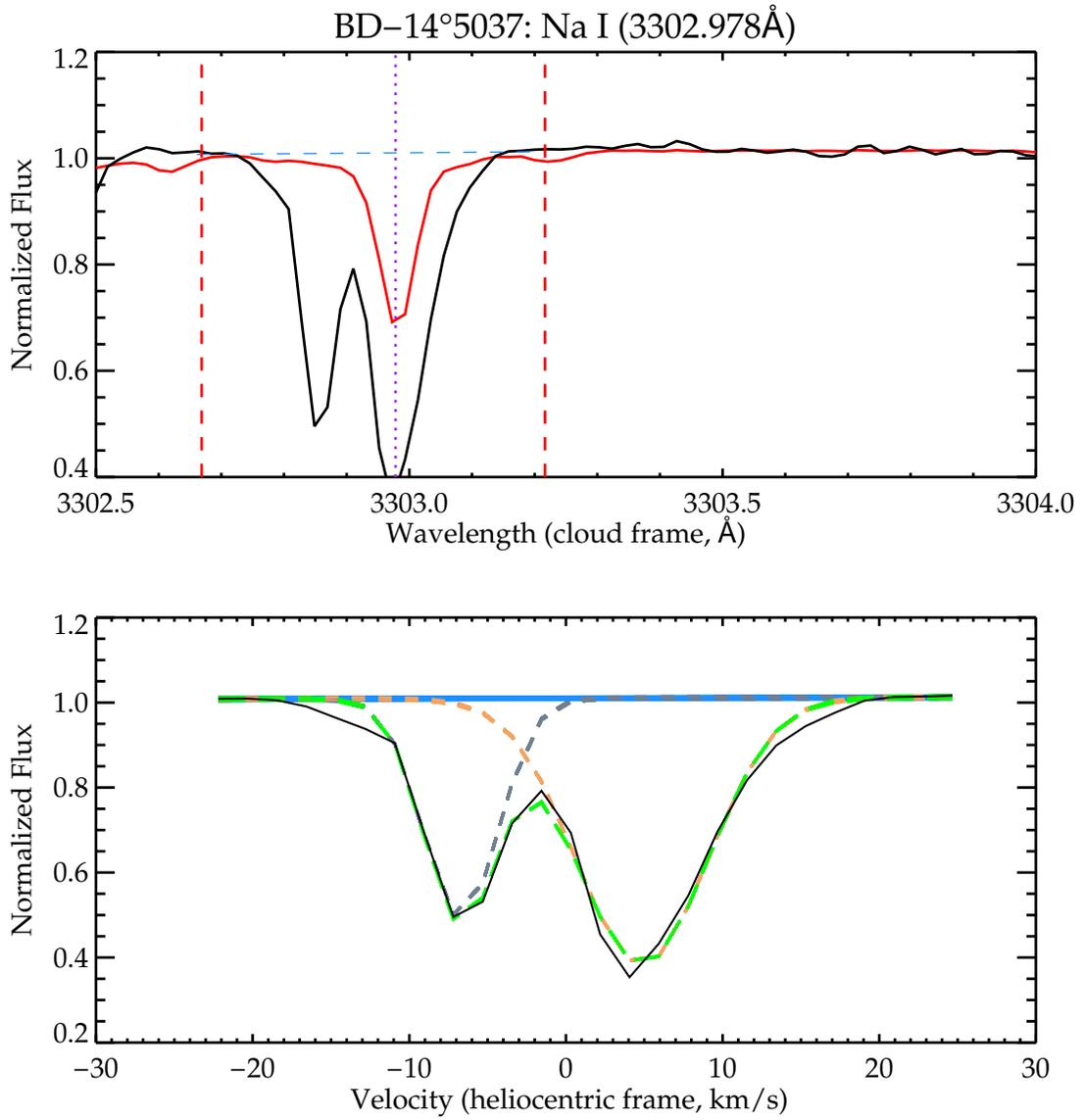}
\caption{\label{Fig:measure_plot}An example measurement of one of the
  \ion{Na}{1} lines in BD-14$^{\circ}$5037 is shown.  The top panel
  shows the target spectrum (black), and the interstellar
  superspectrum of all 51 reddened targets (red), in the interstellar
  rest frame.  The red vertical lines bracket the feature of interest,
  whose rest wavelength is shown with the dotted purple line.  The
  continuum for the feature is shown with a horizontal blue line.  The
  bottom panel depicts the Monte Carlo process on the region of
  interest in heliocentric velocity space of the target spectrum
  (black): the blue horizontal lines are the repositioned continua
  (based on the local S/N); the green swaths are the best fits; and
  the dashed neutral coloured swaths are the Gaussian decompositions.
  { Figures 16.1-16.536 are available in the online version of the
    Journal.  }}
\end{figure}

\end{document}